\documentclass{ieeeaccess}
\makeatletter
\input{t1ptm.fd}
\input{t1phv.fd}
\input{t1pcr.fd}
\makeatother
\DeclareFontShape{T1}{ptm}{n}{it}{<->ssub * ptm/m/it}{}
\DeclareFontShape{T1}{ptm}{n}{sl}{<->ssub * ptm/m/sl}{}
\DeclareFontShape{T1}{ptm}{n}{sc}{<->ssub * ptm/m/sc}{}
\DeclareFontShape{T1}{phv}{n}{it}{<->ssub * phv/m/it}{}
\DeclareFontShape{T1}{phv}{n}{sl}{<->ssub * phv/m/sl}{}
\DeclareFontShape{T1}{pcr}{n}{it}{<->ssub * pcr/m/it}{}
\usepackage{cite}
\usepackage{url}
\usepackage{amsmath,amssymb,amsfonts}
\usepackage{mathrsfs}
\usepackage{algorithm}
\usepackage{algpseudocode}
\usepackage{float}
\usepackage{graphicx}
\usepackage{textcomp}
\usepackage{subcaption}
\usepackage[normalem]{ulem}
\usepackage{listings}
\usepackage{xcolor}

\usepackage[
    colorlinks=true,
    linkcolor=blue,   
    citecolor=blue,   
    urlcolor=blue     
]{hyperref}

\newcommand{\ket}[1]{\left|#1\right\rangle}

\makeatletter
\renewcommand\@makecaption[2]{%
  \vskip\abovecaptionskip
  \sbox\@tempboxa{#1: #2}%
  \ifdim \wd\@tempboxa >\hsize
    #1: #2\par
  \else
    \global\@minipagefalse
    \hb@xt@\hsize{\hfil\box\@tempboxa\hfil}%
  \fi
  \vskip\belowcaptionskip}
\makeatother
\def\BibTeX{{\rm B\kern-.05em{\sc i\kern-.025em b}\kern-.08em
    T\kern-.1667em\lower.7ex\hbox{E}\kern-.125emX}}
\begin{document}
\history{Date of publication xxxx 00, 0000, date of current version xxxx 00, 0000.}
\doi{Unpublished work}

\title{THeatA: Transport- and Heating-aware QEC simulation with Cooling Strategies for QCCD Trapped-Ion Systems}

\author{
\uppercase{Eitaro Ishikawa}\authorrefmark{1}
\uppercase{Noah Hirai}\authorrefmark{1}
\uppercase{Koichiro Miyanishi}\authorrefmark{3}
\uppercase{Hana Ebi}\authorrefmark{1}
\uppercase{Kazufumi Tanji}\authorrefmark{1}
\uppercase{Isamu Kudo}\authorrefmark{1}
\uppercase{Kenji Toyoda}\authorrefmark{6,7}
\uppercase{Masahiro Takeoka}\authorrefmark{1,4}
\uppercase{Hiroki Takahashi}\authorrefmark{5,3}
\uppercase{Takahiko Satoh}\authorrefmark{1}
\uppercase{Shin Nishio}\authorrefmark{1,2}
}

\address[1]{Keio University, Yokohama, Kanagawa, Japan}
\address[2]{University College London, London, United Kingdom}
\address[3]{Qubitcore Inc., Onna-son, Okinawa, Japan}
\address[4]{National Institute of Information and Communications Technology, Tokyo, Japan }
\address[5]{Okinawa Institute of Science and Technology, Onna-son, Okinawa, Japan}
\address[6]{Center for Quantum Information and Quantum Biology, Osaka University,
Osaka, Japan}
\address[7]{Graduate School of Engineering Science, Osaka University,
Osaka, Japan}

\tfootnote{
This project was supported by the JST Moonshot R\&D Grant Number JPMJMS256G.
SN acknowledges support from a JSPS Overseas Research Fellowship.
}

\markboth
{Ishikawa \headeretal: THeatA: Transport- and Heating-aware QEC simulation with Cooling Strategies for QCCD Trapped-Ion Systems}
{Ishikawa \headeretal: THeatA: Transport- and Heating-aware QEC simulation with Cooling Strategies for QCCD Trapped-Ion Systems}

\corresp{Corresponding author: Eitaro Ishikawa (email: \url{ataroh_1031@keio.jp}), Koichiro Miyanishi (email: \url{koichiro.miyanishi@qubitcore.jp}), and Shin Nishio (email: \url{shin.nishio@keio.jp}).}

\begin{abstract}
The quantum charge-coupled device (QCCD) architecture is a promising approach to scalable trapped-ion quantum computing.
It enables flexible qubit interactions by transporting ions between dedicated trap zones.
Native programs for such systems require operations
beyond conventional quantum-circuit primitives,
including ion shuttling, physical SWAPs,
trap splitting and merging, and cooling.
Motional excitation accumulated during transport operations
increases the mean phonon number of ions,
degrading the fidelity of the M{\o}lmer--S{\o}rensen gates
used for two-qubit interactions.
However, existing quantum error correction (QEC) simulation frameworks do not jointly track per-ion motional excitation and explicitly schedule cooling operations with their associated transport time and heating costs.
In this work, we develop a full-stack software framework for QEC experiments and simulations on linear QCCD systems with a single ion species.
The framework comprises 
an intermediate representation (IR) supporting QCCD-native operations,
a compiler that translates syndrome-extraction circuits into the IR,
and a heating-aware noise-injection pipeline that generates noisy Stim circuits from the IR.
To capture the effects of transport and cooling, the compiler updates the mean phonon number of each ion throughout the compiled schedule, and the noise-injection pipeline derives gate-dependent error rates from the tracked phonon states and per-ion timing.
This enables the evaluation of finite-size logical-error scaling
of the rotated surface code
under transport- and heating-aware noise.
Using this framework,
we show that transport-induced heating
prevents logical error rates from decreasing
with increasing code distance
under the evaluated conditions.
We further demonstrate that indirectly cooling the ion pairs involved in two-qubit gates, by merging them with precooled coolant ions immediately before each gate, restores the decrease in logical error rate with code distance.
Finally, by sweeping the duration and post-cooling mean phonon number of an abstract cooling instruction, we characterize the trade-off between these parameters and identify the region in which logical-error scaling is maintained.
 
\end{abstract}


\titlepgskip=-15pt

\maketitle

\clearpage
\section{Introduction}
\label{sec:introduction}

Trapped-ion qubits offer long coherence times and some of the highest gate fidelities demonstrated to date~\cite{Bruzewicz_2019,Harty_2014,Gaebler_2016,smith2025singlequbit,hughes2025smoothgate}. 
The quantum charge-coupled device (QCCD) architecture is a leading approach to scaling this platform~\cite{kielpinski2002architecture,pino2021demonstration}.
A QCCD system divides its trap into multiple zones and physically transports ions between them during a computation.
QCCD-based trapped-ion systems are therefore widely regarded as a promising architecture for realizing large-scale fault-tolerant quantum computing~\cite{jones2025architectingscalabletrappedion, 10.1145/3624062.3624214}.

Fault-tolerant quantum computing (FTQC) relies on quantum error correction (QEC), which repeatedly measures error syndromes to detect and correct errors in encoded logical qubits~\cite{shor1995scheme,Dennis_2002,Fowler_2012}. These measurements require two-qubit gates between data and ancilla qubits. In QCCD architectures, such gates can only be performed between ions in the same zone, requiring ion transport to bring interacting ions together. Consequently, each round of syndrome extraction involves repeated shuttling, splitting, merging, and physical SWAP operations in addition to gates and measurements~\cite{kaushal2019shuttlingbasedtrappedionquantuminformation,hilder2021faulttolerantparityreadoutshuttlingbased}.

Transport is also demanding in its own right.
Each shuttling primitive takes tens to hundreds of microseconds,
so transport rather than gate execution dominates the execution time of shuttling-based demonstrations~\cite{hilder2021faulttolerantparityreadoutshuttlingbased,kaushal2019shuttlingbasedtrappedionquantuminformation}. 
In a linear trap, ions cannot pass one another,
which imposes ordering constraints:
accessing an ion in the interior of a chain
requires splitting the chain
or swapping it with its neighbors.
Because these extra operations depend on
which qubits each gate acts on,
the transport schedule must be planned
jointly with the circuit~\cite{Kreppel_2023,schoenberger2024shuttlingscalabletrappedionquantum}. 

The cost of ion transport is not limited to time.
Shuttling, splitting, merging, and swapping also excite the motional modes of the ions and thereby increase their mean phonon number $\bar{n}$~\cite{kaushal2019shuttlingbasedtrappedionquantuminformation}. 
Although the M{\o}lmer--S{\o}rensen gate~\cite{M_lmer_1999,sorensen2000entanglement}, a standard two-qubit gate in trapped-ion systems, is insensitive to the motional state to first order in the Lamb--Dicke parameter, its residual errors increase with $\bar{n}$~\cite{hirai2026}.
These errors arise from higher-order terms in the Lamb--Dicke expansion and from motional dephasing, which prevents the phase-space trajectory from closing.
Transport-induced heating therefore feeds into the logical error rate.
Moreover, this motional excitation accumulates in each ion, and must therefore be removed by cooling as needed. 
Which ions can be cooled, by which method, and at what timing, however, is strongly constrained by how each cooling method interacts with stored quantum information.

Cooling operations for trapped ions fall into two categories,
distinguished by whether they preserve
the internal quantum state of the ion being cooled.
The first category is direct laser cooling, including Doppler, resolved-sideband, electromagnetically-induced-transparency (EIT), and polarization-gradient (Sisyphus) cooling~\cite{eschner2003laser,diedrich1989zeropoint,roos2000eit,ejtemaee2017sisyphus}.
These cooling methods reach from the Doppler limit down to near the motional ground state,
but they all rely on repeated photon-scattering cycles,
which do not preserve the internal state of the ion being cooled. 
Direct cooling is therefore restricted to ions whose internal state may be discarded, such as ancillae just after measurement.
In the second category,
motional excitation is reduced indirectly
through Coulomb coupling to an auxiliary coolant ion
that can be laser-cooled
without disturbing the quantum information
encoded in the data ion,
as in exchange cooling~\cite{fallek2024exchange}
and sympathetic cooling~\cite{larson1986sympathetic};
with dual-type qubit encodings,
both roles can even be realized
within a single ion species~\cite{yang2022dualtype}. 
The two categories differ in which ions they may act on, in the attainable phonon number, and in their duration.
These constraints make the choice of cooling method,
target, and timing
an important design decision in QCCD systems.

Scaling up a QCCD-based FTQC system
requires a cooling strategy that specifies which ions to cool,
at which location, and at what point in the schedule.
These choices affect both the logical error rate
and the duration of a code cycle, because cooling takes time
while transport keeps adding motional excitation ion by ion.
Their effect therefore depends on the compiled schedule as a whole
rather than on the performance of any single cooling method,
and evaluating a cooling strategy requires circuit-level simulation that tracks transport, heating, and cooling for each ion individually.
Existing QCCD compilers account for transport time but do not explicitly model motional heating~\cite{Kreppel_2023,schoenberger2024shuttlingscalabletrappedionquantum}. Earlier noise-aware evaluations model transport-induced heating in multi-trap systems~\cite{murali2020architectingnoisyintermediatescaletrapped, saki2022muzzle}, but do not evaluate logical error rates. More recent QEC evaluations incorporate heating into logical-error simulations. Jones and Murali~\cite{jones2025architectingscalabletrappedion} use static worst-case phonon estimates and treat cooling as a global noise-model assumption, whereas WISER~\cite{jones2026wiser} tracks phonon occupation as the program executes and applies threshold-triggered, trap-wide sympathetic cooling. However, neither explicitly compiles the transport and cooling operations needed to cool selected data ions while accounting for their time and heating costs. Consequently, the effect of such cooling strategies on logical error rates remains to be evaluated.

\begin{figure}[htbp]
    \centering
    \includegraphics[width=\linewidth]{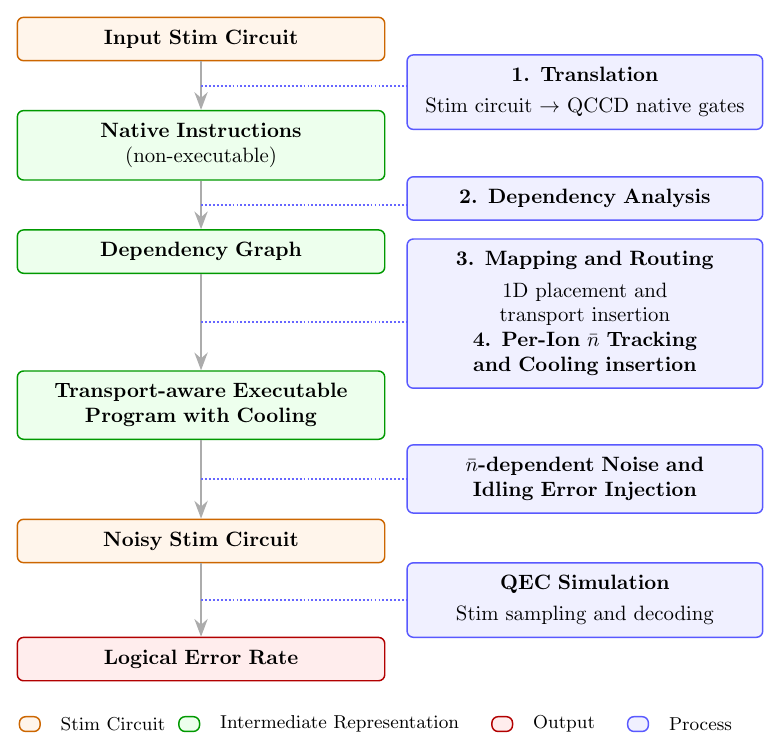}
    \caption{Compilation and evaluation pipeline. An input Stim circuit is translated into
    QCCD native instructions, dependencies are extracted as a DAG, and the
    DAG is mapped onto the 1D trap with transport operations inserted.}
    \label{fig:compiler_workflow}
\end{figure}

In this work, we close this gap
with a transport- and heating-aware
QEC simulation framework
for linear QCCD systems with a single ion species,
as shown in Figure~\ref{fig:compiler_workflow}. 
Its compiler translates syndrome-extraction circuits
into an intermediate representation (IR)
of QCCD-native operations,
covering shuttling, splitting, merging,
physical SWAPs, and cooling. 
Representing cooling at the instruction level,
alongside coolant transport,
is what allows cooling strategies,
the choice of which ions to cool, where, and when,
to be expressed within a compiled schedule.
As the compiler generates the instruction sequence, it tracks the mean phonon number of each ion, accounting for the cumulative effects of transport-induced heating and cooling throughout the schedule.
A noise-injection pipeline then rebuilds
the circuit in Stim form~\cite{Gidney_2021},
derives each gate error from the tracked phonon record and the per-ion timing,
and decodes the sampled circuit with minimum-weight perfect matching decoder~\cite{higgott2022pymatching}
to obtain the logical error rate.
Cooling enters the framework as an explicit instruction
abstracted by a duration and a residual phonon number.
Measured auxiliary qubits are cooled directly
with this instruction.
Data ions are instead cooled indirectly
by repeatedly merging them with laser-cooled ions,
a process we refer to as phonon-averaging cooling,
without simulating the dynamics of
a particular direct cooling protocol. 
The deployment of direct cooling, 
and the merges that carry its effect to the data ions,
thereby become compilation choices
that the framework compares on an equal footing.

Our simulations yield three findings.
First, without any cooling, enlarging the code makes the logical qubit worse. We judge improvement of logical error rate $p_L$ from distance $d$ to $d+2$ by the ratio
$\Lambda_{d,d+2} = p_L(d) / p_L(d+2)$,
and count scaling as established
only when the 95\% interval of the measured ratio
lies above one. 
Without cooling, the measured ratio drops to $\Lambda_{3,5} = 0.08$, because longer transport schedules heat the ions
faster than the added redundancy protects them.
Replaying the same compiled schedules with heating removed
restores scaling
($\Lambda_{3,5} = 4.7$ and $\Lambda_{5,7} = 9.6$),
so under the modeled parameters the entire loss of scaling is heating-induced.
Second, within the evaluated model,
the choice of cooling strategy determines
whether the logical error rate decreases
with the code distance.
The simplest strategy,
cooling the ancillae after each measurement,
lowers the logical error rate at every distance,
but the rate still fails to decrease with the code distance.
Scaling at both distance pairs is achieved only when
phonon-averaging cooling is applied to the data ions
immediately before each two-qubit gate,
keeping their phonon numbers below a threshold
($\Lambda_{3,5} = 3.99$ and $\Lambda_{5,7} = 5.39$,
with their 95\% intervals above one). 
Third, sweeping the duration and the residual phonon number
of the abstract cooling instruction
maps the cooling requirements for logical-error scaling.
When phonon-averaging cooling is applied
with durations of one to ten milliseconds
and residual phonon numbers up to two,
scaling is maintained in nearly all cases.
The sweep in the duration--residual plane further shows
that the two parameters trade off against each other,
and that scaling is lost
at durations between $50$ and $75\,\mathrm{ms}$
even with zero residual phonon number,
because the longer schedule lets background heating
raise the phonon numbers at the gates
far above the residual value.
These requirements concern the parameters of
the abstract cooling instruction,
rather than the performance of
any particular physical cooling method. 

In summary, this paper makes the following contributions:
\begin{itemize}
    \item A transport- and heating-aware compiler and QEC simulation framework for QCCD systems.
    It tracks phonon accumulation for each ion through the full instruction sequence, exposes cooling as an explicit instruction sequence, and evaluates logical error rates end-to-end.
    \item A systematic evaluation of cooling strategies
    based on phonon-averaging cooling with direct cooling of coolant ions,
    spanning the placement of coolants
    and the scheduling of cooling with transportation. 
    \item An evaluation of the cooling requirements
    for finite-size logical-error suppression against the code distance,
    expressed as its duration and the residual phonon number
    of an abstract direct cooling instruction.
\end{itemize}
The source code of the proposed compiler and simulation framework is publicly available on GitHub~\cite{theata_github}.

The remainder of this paper is organized as follows.
Section~\ref{sec:preliminaries} introduces the 1D QCCD architecture, its native operations, and the cooling methods our models build on.
Section~\ref{sec:related} positions this work
against existing compilers and noise-aware evaluations
for multi-trap trapped-ion systems.
In Section~\ref{sec:method1} we propose a compiler that maps syndrome-extraction circuits onto the QCCD architecture by inserting the required ion-transport operations.
We also introduce a simulation pipeline for evaluating the logical error rate based on these hardware-aware circuits.
Section~\ref{sec:experiments} evaluates cooling strategies
and measures the cooling requirements
for logical-error scaling.
Section~\ref{sec:conclusion} summarizes the findings
and discusses limitations and future work.
\begin{figure*}[htbp]
    \centering
    \includegraphics[width=\linewidth]{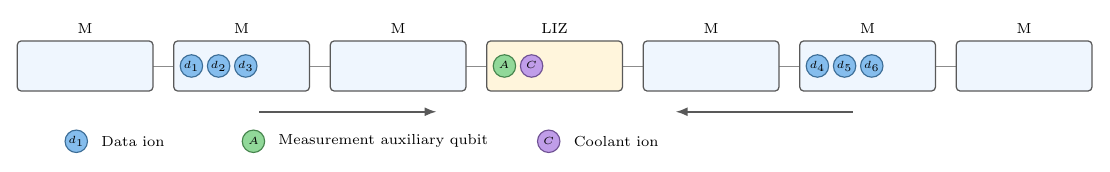}
    \caption{A schematic diagram of the 1D QCCD architecture consisting of a laser interaction zone (LIZ) and memory zones (M). Unitary gates, measurements, resets, and cooling operations are performed in the LIZ, requiring the participating ions to be shuttled from the memory zones to the LIZ.}
    \label{fig:zone}
\end{figure*}

\section{Preliminaries}
\label{sec:preliminaries} 

\subsection{Stabilizer Codes and rotated Surface Code}
\label{subsec:stabilizer}

A $[\![n, k, d]\!]$ stabilizer code encodes $k$ logical qubits
into $n$ physical qubits with minimum distance $d$,
defining the code space as the joint $+1$ eigenspace
of a set of mutually commuting Pauli operators $\{S_i\}$,
called stabilizer generators~\cite{gottesman1997stabilizercodesquantumerror, nielsen2010quantum}.
The rotated surface code~\cite{Tomita_2014} places data qubits on a two-dimensional lattice and employs weight-$4$ stabilizer generators in the bulk and weight-$2$ generators at the boundaries for both $X$- and $Z$-type stabilizers. Surface codes are promising candidates for fault-tolerant quantum computing due to their high error thresholds~\cite{Dennis_2002,Fowler_2012}. We use the rotated surface code to evaluate our compiler and cooling strategies.

\subsection{Syndrome Measurement Circuits}
\label{subsec:syndrome}
To check whether an encoded state lies in the code space, each stabilizer generator is measured indirectly through measuring an ancilla qubit. For a weight-$4$ $Z$-type generator, the ancilla is
initialized to $\ket{0}$, four CNOT gates transfer the parity onto the
ancilla, and the ancilla is measured in the computational basis; for an
$X$-type generator, Hadamard gates sandwich the CNOT sequence so that the
effective measurement is in the $X$ basis~\cite{Fowler_2012}.

\subsection{1D QCCD Architecture}
\label{subsec:qccd}

Trapped-ion quantum computers realize qubits as internal states of atomic
ions confined by electromagnetic fields~\cite{HAFFNER_2008}. The
Quantum Charge-Coupled Device (QCCD) architecture~\cite{kielpinski2002architecture}
addresses the scalability limitation of a single linear trap by dividing the
trap region into functional \emph{zones}, connected by segmented DC
electrodes that enable ion transport between them.
These zones are generally classified into
\emph{laser interaction zones~(LIZs)},
where gate operations, initializations, cooling, and measurements are performed,
and \emph{memory zones}, which store ions.
In this work, we consider a linear configuration
of seven zones with a single LIZ at the center,
denoted \texttt{M-M-M-L-M-M-M} as shown in Figure~\ref{fig:zone}.
Because ions in the same zone form a linear chain,
they cannot pass each other during transport,
and changing their order requires
physical SWAP operations. 

\subsubsection{Native Operations} 
\label{subsec:native_ops}

In this paper, we adopt the following set of native operations,
grouped into two categories.

\paragraph{Computational operations}
These change the quantum state and are executed inside the LIZ:
\begin{itemize}
    \item \textbf{Single-qubit rotation} $R(\theta, \phi)$, a rotation by
    angle $\theta$ around an axis in the $xy$-plane of the Bloch sphere
    defined by phase $\phi$~\cite{HAFFNER_2008, Maslov_2017}.
    \item \textbf{M{\o}lmer--S{\o}rensen (MS) gate}\\
    $U_{MS}(\theta)=\exp(-i\tfrac{\theta}{2}\sigma_x^{(i)}\sigma_x^{(j)})$,
    the two-qubit entangling gate adopted in this work~\cite{M_lmer_1999, Maslov_2017}.
    \item \textbf{State preparation} (\textsc{Reset}) and \textbf{measurement}
    in the computational basis.
\end{itemize}

\paragraph{Transport operations}
These reconfigure ion positions without changing the internal state:
\begin{itemize}
    \item \textbf{Shuttling}: adiabatically transport an ion or crystal to
    another zone.
    \item \textbf{Split / Merge}: separate a crystal into two or combine two
    crystals into one.
    \item \textbf{Physical SWAP}: reorder two ions within the same
    potential well. 
\end{itemize}

\subsubsection{Motional Heating and Gate Fidelity} 
\label{subsec:heating}

The ions in the same potential share quantized collective motional modes,
conventionally divided into \emph{axial} modes along the trap axis and
\emph{transverse} modes perpendicular to the trap axis~\cite{HAFFNER_2008, Bruzewicz_2019}.
While qubits are encoded in internal states of the ions,
two-qubit gates such as the MS gate
are mediated by a shared motional mode,
which couples the internal states of the participating
ions~\cite{M_lmer_1999, HAFFNER_2008}.

Transport operations disturb this motion. Shuttling, splitting, merging, and physical SWAPs all transfer energy into the motional modes and increase the mean phonon number 
$\bar{n}$~\cite{kielpinski2002architecture, Poschinger_2009, kaushal2019shuttlingbasedtrappedionquantuminformation}.
Our model tracks both transport-induced heating
and background heating,
the latter proportional to elapsed time,
together with their removal by dedicated cooling operations.
Both heating and cooling are accumulated separately
for each ion and each motional mode. 

The fidelity of the MS gate degrades as $\bar{n}$ increases~\cite{hirai2026} although the gate is insensitive to the motional state to first order~\cite{M_lmer_1999},
residual higher-order couplings reduce the achievable fidelity for hotter ions~\cite{Ballance_2016, Gaebler_2016}.
Cooling operations can restore $\bar{n}$ to a low residual value at the cost of additional execution time.
We use the dependence of MS-gate fidelity on $\bar{n}$ to map motional excitation to gate error rates, with a lookup table derived from a physical simulation of the MS gate in Ref.~\cite{hirai2026}.

\subsection{Cooling Methods for Trapped-Ion Qubits}
\label{subsec:cooling}

Transport-induced motional excitation motivates introducing cooling operations during the computation. Cooling methods can be classified by whether they preserve the internal quantum state
of the ion being cooled.

\paragraph{Direct laser cooling}
Doppler cooling~\cite{wineland1987lasercooling,eschner2003laser}, resolved-sideband cooling~\cite{diedrich1989zeropoint,roos1999sideband}, and electromagnetically-induced-transparency (EIT) cooling~\cite{roos2000eit} reduce motional excitation through laser-driven transitions with different attainable motional temperatures and cooling durations.
Because they involve photon scattering,
they do not preserve
arbitrary quantum information
encoded in the illuminated internal states.
In our model,
direct laser cooling is therefore restricted to
ancillae after measurement
and dedicated coolant ions.

\paragraph{State-preserving indirect cooling}
Data ions must preserve their quantum information
during syndrome extraction.
Their motional excitation can instead be reduced
through coupling to auxiliary ions
that can be laser-cooled
without directly cooling the data qubits in LIZ.
In exchange cooling,
an initially cooled ion is brought close to
a data ion in a separate potential well,
and resonant motional coupling
transfers excitation between them~\cite{fallek2024exchange}.
In sympathetic cooling,
data and coolant ions share collective motional modes,
allowing laser cooling of the coolant
while preserving the data ion's internal
state~\cite{larson1986sympathetic,barrett2003sympathetic,home2009memory}.
State-selective cooling using protected internal levels
also permits implementations
with ions of the same
species~\cite{allcock2021omg,yang2022dualtype}.

\paragraph{Cooling model used in this work}
Our simulator represents direct laser cooling
by a \texttt{Cool} instruction and it has two parameters.
They are the post-cooling residual mean phonon number after direct laser cooling $\bar{n}_{\mathrm{res}}$
and the instruction duration $t_{\mathrm{cool}}$.
Individual laser-cooling stages and their physical dynamics are not simulated.
All ions in the framework are modeled as a single species with common motional and timing parameters.

Merge operations follow an idealized averaging model, 
regardless of which ions are involved.
When a chain $C_1$ of $n_1$ ions is merged
with a chain $C_2$ of $n_2$ ions,
the mean phonon number of the resulting chain
is updated for each tracked mode
$\mu \in \{\mathrm{ax}, \mathrm{tr}_1, \mathrm{tr}_2\}$ as
\begin{equation}
    \bar{n}_{C_1\cup C_2,\mu}
    = \frac{n_1\bar{n}_{C_1,\mu}
            + n_2\bar{n}_{C_2,\mu}}
           {n_1+n_2}
      + e_{\mu}^{\mathrm{merge}},
    \label{eq:phonon_averaging}
\end{equation} 
where $\bar{n}_{C_i,\mu}$ is the mean phonon number
assigned to the ions of $C_i$ before the merge,
and $e_{\mu}^{\mathrm{merge}}$
is the merge-induced increase in the tracked mean phonon number.
For simplicity, we do not resolve the individual radial normal modes
and instead assume that the phonons are distributed equally
among all of them.
The resulting value is assigned to every ion
in the merged chain.
We refer to this size-weighted equalization step
as \emph{phonon averaging};
it redistributes the modeled motional excitation
without removing it,
and the merge-heating increment is added afterward.

By repeatedly merging data ions with precooled
auxiliary ions, separating the auxiliary ions,
and laser-cooling them, the mean phonon number
of the data ions can be reduced.
We refer to this idealized procedure as
\emph{phonon-averaging cooling}.
Unlike experimentally demonstrated exchange cooling,
this model assumes complete equalization upon merging
and does not resolve the finite-time dynamics
of the energy exchange.
The implications of this approximation
are discussed in Section~\ref{sec:conclusion}.
\section{Related Work} 
\label{sec:related}
Prior work relevant to this paper falls into three lines of work:
transport-aware compilers for NISQ devices~\cite{10.1145/3624062.3624214, Kreppel_2023, schoenberger2024shuttlingscalabletrappedionquantum},
noise-aware NISQ evaluations of multi-trap systems~\cite{murali2020architectingnoisyintermediatescaletrapped, saki2022muzzle},
and a transport- and heating-aware QEC simulation workflow~\cite{jones2025architectingscalabletrappedion, jones2026wiser}.
Table~\ref{tab:research_position} summarizes the comparison.

The first line treats transport as time or as operation counts without explicitly simulating gate fidelity depending on transport-induced motional heating.
TISCC~\cite{10.1145/3624062.3624214}
generates surface-code patch operations
for grid-based QCCD hardware,
Kreppel et al.~\cite{Kreppel_2023}
reduce gate counts,
thereby shortening the shuttling schedule,
and Schoenberger et al.~\cite{schoenberger2024shuttlingscalabletrappedionquantum}
schedule shuttling in time steps.
The heating itself appears in
neither the cost model nor the evaluation.

The second line evaluates multi-trap systems
under noise on NISQ workloads.
The simulator of Murali et al.~\cite{murali2020architectingnoisyintermediatescaletrapped}
models gate fidelity as degrading with the motional energy an ion chain accumulates, and Saki et al.~\cite{saki2022muzzle}
reduce shuttle counts under the same model.
Here motional energy is a single scalar per chain, cooling is not supported,
and neither work evaluates a logical error rate.

Jones and Murali~\cite{jones2025architectingscalabletrappedion}
evaluate compiled surface-code schedules
using a heating-aware error model with Stim,
utilizing a fixed worst-case phonon number
per transport primitive and cooling as a global noise-model variant.
In our compiler, heating accumulates per ion,
and cooling is an instruction
placed by a trigger policy
(Section~\ref{subsec:phonon_tracking}).
Their traps offer all-to-all connectivity internally,
with reordering implemented as a three-gate logical SWAP;
our ions cannot bypass one another,
so reordering takes time and adds heat, as details are described in Section~\ref{subsec:qccd}.
Concurrent work from the same group,
WISER~\cite{jones2026wiser},
narrows this gap:
phonon occupation is updated as the program executes,
a single threshold shared by all traps triggers recooling,
and cooling resets the heating of whole target traps.
Our tracking is per ion and mode-resolved,
our coolant ions are placed and transported
by the schedule,
and zone-local control opens the strategy space
that Section~\ref{sec:experiments} explores.

\begin{table*}[htbp]
    \centering
    \caption{Research position of this work relative to noise-aware evaluations of multi-trap trapped-ion systems.}
    \label{tab:research_position}
    \renewcommand{\arraystretch}{1.3}
    \begin{tabular}{l|p{0.185\textwidth}|p{0.185\textwidth}|p{0.185\textwidth}|p{0.185\textwidth}}
        \hline
        & \textbf{NISQ multi-trap}~\cite{murali2020architectingnoisyintermediatescaletrapped, saki2022muzzle}
        & \textbf{Jones \& Murali}~\cite{jones2025architectingscalabletrappedion}
        & \textbf{WISER}~\cite{jones2026wiser}
        & \textbf{This work} \\
        \hline
        \hline
        Transport cost model
        & Path-dependent time and motional heating~\cite{murali2020architectingnoisyintermediatescaletrapped, saki2022muzzle}
        & Per-operation transport time and motional heating
        & Per-operation transport time and motional heating 
        & Distance-dependent time and per-ion \\
        \hline
        Phonon state
        & Dynamic, one scalar per chain
        & Static, per-operation worst-case estimates
        & Dynamic, per-trap
        & Dynamic, per-ion, mode-resolved \\ 
        \hline
        Cooling strategy
        & None
        & Global cooling assumption in the noise model
        & Global phonon threshold, trap-wide sympathetic cooling with scattering penalty 
        & Direct cooling of coolant and auxiliary ions, indirect cooling on data ions accounting for transport time and heating
        \\
        \hline
        Trap / zone model
        & Uniform gate-capable traps, all-to-all within a trap
        & Uniform gate-capable traps, all-to-all within a trap
        & Uniform gate-capable traps, all-to-all within traps, shared transport and gate control
        & 1D ordered chain, memory--LIZ separation, no-bypass constraint \\
        \hline
        Logical-error evaluation
        & None~\cite{murali2020architectingnoisyintermediatescaletrapped}
        
        Program fidelity~\cite{saki2022muzzle}
        & Stim
        & Stim
        & Stim \\
        \hline
    \end{tabular}
\end{table*}

To our knowledge, no prior evaluation combines per-ion, mode-resolved heating across transport operations with selective cooling that respects the restrictions on direct data-ion cooling. Our framework integrates these capabilities into a compiler for a memory-LIZ-separated 1D QCCD architecture, enabling us to evaluate how cooling strategies affect logical-error scaling.
\section{Compiler and Noise Simulation Pipeline}
\label{sec:method1}

This section describes the proposed compiler and the evaluation pipeline built on top of it.
The compiler translates a syndrome measurement circuit specified in the Stim format~\cite{Gidney_2021} into a physically executable instruction sequence for a 1D QCCD architecture as described in Section~\ref{subsec:translation} - \ref{subsec:routing}.
Three components then turn the compiled sequence into evaluable quantities.
An execution-time estimator described in Section~\ref{subsec:estimator},
a per-ion phonon tracker in Section~\ref{subsec:phonon_tracking},
and a noise-injection and decoding pipeline that produces a logical error rate in Section~\ref{subsec:noise_pipeline}.

The compilation pipeline as shown in Figure~\ref{fig:compiler_workflow} consists of three stages:
\begin{enumerate}
    \item \textbf{Translation:} the input Stim circuit is translated into the
    QCCD native gate set introduced in Section~\ref{subsec:qccd},
    ignoring physical constraints such as ion positions or transport delays.
    \item \textbf{Dependency Analysis:} dependencies between instructions
    are analyzed to build a directed acyclic graph (DAG) that captures the
    qubit-level execution order.
    \item \textbf{Mapping and Routing:} the compiler maps the DAG onto the
    physical 1D trap by placing ions and inserting transport operations
    (\texttt{Shuttle}, \texttt{Split}, \texttt{Merge}, \texttt{SWAP})
    so that all hardware constraints are satisfied.
    \item \textbf{$\bar{n}$ tracking and cooling insertion:} track the motional heating per ion and insert direct cooling to ancillae and coolants, and indirect cooling to data ions with additional transport operations by phonon averaging. 
\end{enumerate}

\subsection{Translation to Intermediate Representation}
\label{subsec:translation}

The input Stim circuit is converted into an intermediate representation with native gates. 
We define a domain-specific language (DSL) for the intermediate representation which supports primitives for QCCD-specific instructions. This is defined in extended Backus--Naur Form~\cite{iso14977, Scowen1993GenericBS} and we parse it with Lark~\cite{lark} to produce an abstract syntax tree (AST). 
The full grammar and the instruction set are given in Appendix~\ref{app:ir}.

\subsection{Dependency Analysis}
\label{subsec:dag}

After parsing, the compiler builds a directed acyclic graph (DAG)
$G=(V,E)$ in which each vertex 
represents a \emph{computational} instruction (\texttt{R}, \texttt{MS},
\texttt{Reset}, \texttt{Measure}) and each directed edge represents a
qubit-level dependency,
so that the reachability of the graph defines the partial order
that execution must respect.
Explicit \texttt{Cool} and \texttt{Barrier} instructions,
when present in the input program, also can be in the DAG.
Because a \texttt{Cool} acts on a whole zone
whose occupants are not determined until routing,
it is ordered conservatively as a synchronization node
that follows every preceding instruction.
The DAG is built in a single pass over the parsed
instruction sequence:
each instruction is connected to the most recent
earlier instruction acting on each of its qubits
(Algorithm~\ref{alg:dag_construction} in Appendix~\ref{app:ir}).

\subsection{Mapping and Routing}
\label{subsec:routing}

The mapping and routing stage converts the dependency graph into an executable instruction sequence for the 1D QCCD architecture. Starting from the initial ion placement such as \emph{boustrophedon placement} described in Appendix \ref{app:qubit-mapping}, the compiler selects dependency-ready instructions and determines the transport operations required to execute them under the physical constraints of the architecture. It inserts Split, Shuttle, Merge, and physical SWAP operations as needed, while updating the positions and chain configurations of the ions. The resulting instruction sequence preserves the dependencies of the input circuit and satisfies the operational constraints of the 1D QCCD architecture described below.

The mapping and routing stage must satisfy the following operational constraints:
\begin{enumerate}
    \item \textbf{1D Transport (No-Bypass):}
    Ions cannot bypass each other during transport.
    Reordering requires an explicit \texttt{Swap}.

    \item \textbf{LIZ Constraint:}
    Computational operations can only be executed in a LIZ.
    For two-qubit gates, both target ions must reside in the same LIZ.

    \item \textbf{Zone Isolation:}
    \texttt{Measure} and \texttt{Reset} act on all ions
    in the target zone. Therefore, spectator ions whose
    quantum states must be preserved must be evacuated
    before these operations.

    \item \textbf{Resource Exclusivity:}
    Each LIZ is assumed to have a single set of control lasers,
    and operations within the same zone are executed serially.

    \item \textbf{Zone Capacity:}
    The number of ions in a zone must not exceed its capacity during transport. For example, a \texttt{Shuttle} that would exceed the zone capacity is not permitted.
\end{enumerate}

The routing procedure processes the dependency graph by repeatedly selecting a dependency-ready instruction. Among these instructions, the compiler prioritizes those whose operands are already in the required positions and breaks ties by their order in the input circuit. This selection rule makes the execution order deterministic for a given input circuit and initial ion placement.

For each selected instruction, the compiler checks whether it can be executed under the operational constraints
defined above. If the current ion configuration does not satisfy these constraints, the compiler inserts the necessary transport operations before issuing the instruction.
Throughout this process, the compiler updates the positions and chain configurations of the ions after each transport operation. This allows subsequent routing decisions to reflect the physical configuration produced by preceding instructions.

The following procedures describe how the compiler resolves the conflicts under the current ion configuration associated with gate execution, zone isolation, and ion reordering.

\subsubsection*{Transport for Gate Execution}

When the target ions of a computational gate are not
in the required LIZ, the compiler transports them
to the interaction zone. The target ions are isolated
from their current chains using \texttt{Split},
moved through the trap using \texttt{Shuttle},
and combined in the LIZ using \texttt{Merge}.
If an intervening ion prevents the required transport,
the compiler inserts a physical \texttt{Swap}
to resolve the ordering conflict.

\subsubsection*{Zone Isolation for Measurement and Reset}

Before executing \texttt{Measure} or \texttt{Reset},
the compiler evacuates spectator ions whose quantum
states must be preserved from the target zone.
These ions are isolated using \texttt{Split},
transported using \texttt{Shuttle}, and merged
into the nearest memory zone with available capacity.
Coolant ions are allowed to remain in the target zone.

\subsubsection*{Resolving Linear Constraints (Swap)}

In the 1D QCCD architecture, ions cannot bypass
one another during transport. Reordering therefore
requires a physical \texttt{Swap}.
Because this operation acts on an isolated pair,
the required transport sequence depends on the
position of the target pair within its chain.

\begin{enumerate}
    \item \textbf{Isolated pair (chain length = 2):}
    The pair is already isolated, so only
    \texttt{Swap} is required.

    \item \textbf{Edge pair (chain length $\geq 3$):}
    A single \texttt{Split} isolates the target pair.
    The sequence is
    \texttt{Split} $\rightarrow$ \texttt{Swap}
    $\rightarrow$ \texttt{Merge}.

    \item \textbf{Interior pair (chain length $\geq 4$):}
    Two \texttt{Split} operations isolate the pair
    from both sides. The sequence is
    \texttt{Split} $\times 2$
    $\rightarrow$ \texttt{Swap}
    $\rightarrow$ \texttt{Merge} $\times 2$.
\end{enumerate}

The routing algorithm described above is deterministic and greedy. It follows the instruction-selection rule described above and resolves physical conflicts as they arise, without globally optimizing the transport schedule. Consequently, the generated instruction sequence is not guaranteed to minimize execution time. If routing cannot proceed, the compiler terminates with a diagnostic rather than producing an executable instruction sequence.

\subsection{Execution Time Estimation}
\label{subsec:estimator}

Because each zone executes its operations serially,
the total execution time of a compiled sequence can be modeled
as the sum of per-instruction durations along the critical path.
The estimator takes the routed instruction sequence as input
and returns this total duration.

Table~\ref{tab:timing_parameters} lists the duration assigned
to each native instruction.
These values are taken from trapped-ion experimental
literature~\cite{hilder2021faulttolerantparityreadoutshuttlingbased, Poschinger_2009, kaushal2019shuttlingbasedtrappedionquantuminformation}.\footnote{They are platform-specific and depend on physical details such as electrode pitch,
DAC update rate, and trap geometry.}

\begin{table}[H]
    \centering
    \caption{Per-instruction durations for the execution time estimator.\\}
    \label{tab:timing_parameters}
    \renewcommand{\arraystretch}{1.2}
    \begin{tabular}{l|l|r|l}
        \hline
        \textbf{Instruction} & \textbf{Symbol} & \textbf{Duration} & \textbf{Source} \\
        \hline
        \hline
        Single-qubit gate (\texttt{R}) & $t_{1q}$ & $4\ \mu$s & \cite{hilder2021faulttolerantparityreadoutshuttlingbased} \\
        Two-qubit gate (\texttt{MS}) & $t_{2q}$ & $59\ \mu$s & \cite{Poschinger_2009} \\
        Measurement (\texttt{Measure}) & $t_{\mathrm{meas}}$ & $1000\ \mu$s & \cite{hilder2021faulttolerantparityreadoutshuttlingbased} \\
        Reset (\texttt{Reset}) & $t_{\mathrm{reset}}$ & $41\ \mu$s & Est.\ from \cite{hilder2021faulttolerantparityreadoutshuttlingbased} \\
        \hline
        Split (\texttt{Split}) & $t_{\mathrm{split}}$ & $100\ \mu$s & \cite{kaushal2019shuttlingbasedtrappedionquantuminformation} \\
        Merge (\texttt{Merge}) & $t_{\mathrm{merge}}$ & $100\ \mu$s & \cite{kaushal2019shuttlingbasedtrappedionquantuminformation} \\
        Swap (\texttt{Swap}) & $t_{\mathrm{swap}}$ & $20\ \mu$s & \cite{kaushal2019shuttlingbasedtrappedionquantuminformation} \\
        \hline
        Shuttle & $t_{\mathrm{shuttle}}$ & $9\ \mu$m$/\mu$s & \cite{kaushal2019shuttlingbasedtrappedionquantuminformation} \\
        \hline
        Cooling & $t_{\mathrm{cool}}$ & - & \\
        \hline
        
    \end{tabular}
\end{table}

The duration of \texttt{Shuttle} is not a constant,
which scales with the distance traveled.
Assuming a constant-velocity profile, we compute
\begin{equation}
    t_{\mathrm{shuttle}} = \frac{L_{\mathrm{zone}} - n \cdot d_{\mathrm{site}}}{v_{\mathrm{shuttle}}},
\end{equation}
where $L_{\mathrm{zone}}$ is the inter-zone transport distance,
$d_{\mathrm{site}}$ is an effective per-ion contribution
to the remaining transport distance,
and $n$ is the number of ions already present in the destination zone.
We adopt $L_{\mathrm{zone}} = 1000\ \mu\mathrm{m}$
and $d_{\mathrm{site}} = 5\ \mu\mathrm{m}$ in this paper.

\subsection{Per-Ion Phonon Tracking}
\label{subsec:phonon_tracking}

Every transport primitive heats the ions that move.
To expose this cost to the evaluation, the routing stage tracks a motional state for every ion alongside its position.
We track a mean phonon number for one axial and two transverse modes, initialized to zero. Although the values are stored separately for each ion, all ions belonging to the same ion chain are assigned identical mean phonon numbers in each mode in our model.

Transport operations update these accumulators.
\texttt{Split} and \texttt{Swap}
add fixed, mode-specific amounts
to the ions they act on.
\texttt{Shuttle} adds an axial contribution proportional to the transport distance, using the same distance model as the execution-time estimator
in Section~\ref{subsec:estimator}, together with a fixed transverse contribution.

When two ion chains are merged, the routing stage updates their tracked mean phonon numbers according to Eq.~\eqref{eq:phonon_averaging}.
This update applies phonon averaging followed by the merge-induced heating increment, and the resulting value for each mode is assigned to every ion in the merged chain.

A dedicated cooling instruction,
\texttt{Cool},
resets the tracked mean phonon number of the ion chain in a target zone.
It sets every tracked accumulator of every ion in the chain to configurable residual values $\bar{n}_{\mathrm{res}}$, at the execution-time cost $t_{\mathrm{cool}}$. 
When, where, and on which ions this instruction is issued
is a policy decision of the compiler. We define this cooling policy in the following subsection. 

\begin{figure*}[htbp]
    \centering
    \includegraphics[width=\linewidth]{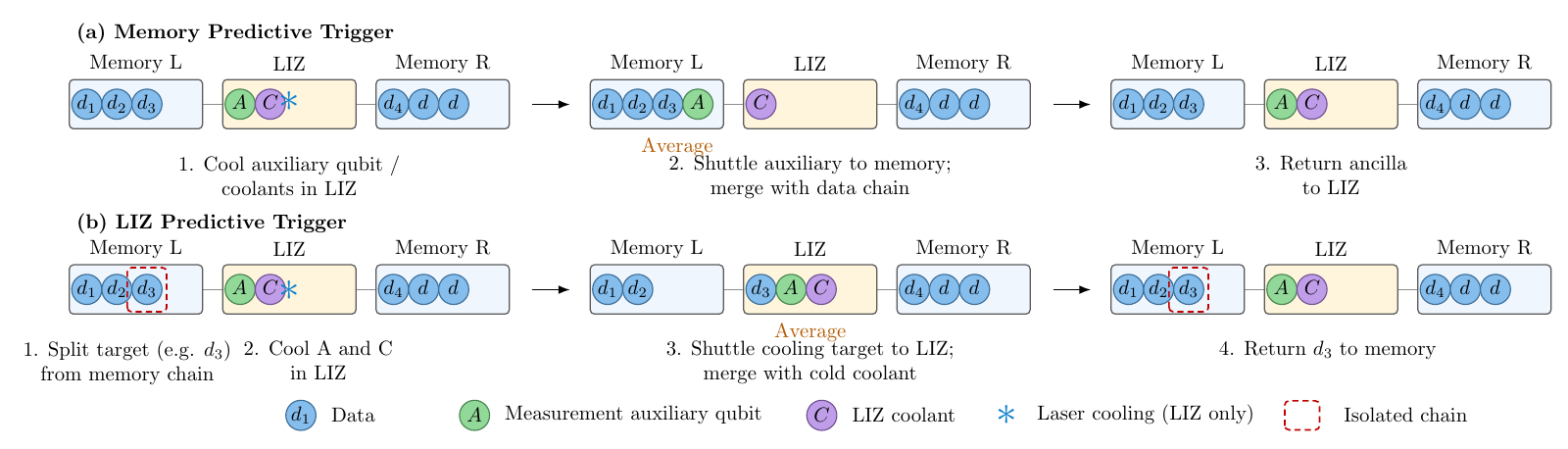}
    \caption{
    Cooling strategies based on predicted phonon numbers. These cooling cycles are repeated up to $N_{\mathrm{max}}$ times. (a) Memory predictive trigger: When the predicted mean phonon number of a target ion exceeds the threshold $\theta$, a precooled auxiliary ion is shuttled from the LIZ to the memory zone and merged with the data-ion chain to reduce its motional excitation. (b) LIZ predictive trigger: The target data ion is shuttled to the LIZ and indirectly cooled through merging with resident coolant ions. Before transport, the target ion is isolated from the other data ions by splitting the chain in the memory zone, preventing its reduced phonon number from being averaged with that of the remaining memory ions upon return.}
    \label{fig:trigger_cooling}
\end{figure*}

\subsection{Cooling Policies}
\label{subsec:cooling_policies}
We implement cooling policies that determine when and where cooling operations are inserted into the compiled instruction sequence.
As described in Section~\ref{sec:preliminaries}, direct cooling is restricted to coolant ions and ancillae that are not actively involved in a stabilizer measurement, while data ions are cooled indirectly through phonon averaging.

All policies include baseline \emph{LIZ cooling}, performed after each stabilizer measurement. At this point,
the LIZ contains the already measured ancilla and, when permanent LIZ coolants are enabled, two resident coolant ions. 
Cooling these ions reduces the motional excitation of the LIZ chain before subsequent gate operations.

In addition to baseline LIZ cooling, we consider two cooling policies that use phonon-number
thresholds to determine when cooling is needed, but differ in where and when the cooling procedure
is initiated as shown in Fig.~\ref{fig:trigger_cooling}.

\paragraph*{Policy 1: Memory Predictive Trigger}
After each stabilizer measurement, the compiler predicts the phonon numbers of the data ions participating in the next stabilizer measurement.
Each prediction is obtained by adding a measurement-window increment, estimated in a preparatory compilation pass, to the ion's currently tracked phonon number.

If any predicted value exceeds the threshold $\theta$, the cooled ancilla is shuttled from the LIZ to the corresponding memory chain and merged with it to reduce its motional excitation. Each visit removes at most a chain-length fraction. The ancilla is then returned to the LIZ. This cooling cycle is repeated while the threshold is exceeded, up to a maximum of $N_{\max}$ cycles per checkpoint.

\paragraph*{Policy 2: LIZ Predictive Trigger}
Immediately before each two-qubit gate, the compiler checks the tracked phonon number of the participating data ion, after it has entered the LIZ and has already been diluted once by merging with the resident chain on arrival.

If the phonon number exceeds the threshold $\theta$, the data--ancilla pair is isolated from the resident
coolant ions. The LIZ are then laser-cooled, and the pair is merged back with them to reduce its motional excitation through phonon averaging. This cycle is repeated until the tracked phonon number falls to or below $\theta$, the reduction in a completed cycle is less than $10^{-3}$ quanta, or the maximum of $N_{\max}$ cycles is reached.

For both predictive policies, the threshold $\theta$ determines when cooling is triggered but does not guarantee an upper bound on the phonon number at gate execution, owing to subsequent heating and finite cooling cycles.

\subsection{Noise Injection and Logical Error Rate Estimation}
\label{subsec:noise_pipeline}

In this last stage, the compiled instruction sequence is converted into a noisy Stim circuit to estimate the logical error rate. The evaluation pipeline consists of three steps:
(1) reconstructing the logical circuit from the compiled instructions, (2) injecting noise channels based on the tracked phonon numbers and per-ion execution timelines, and (3) estimating the logical error rate through circuit sampling and decoding.

Unlike conventional circuit-level noise models with fixed error rates, our pipeline assigns error channels to individual operations based on the compiled physical schedule. In particular, two-qubit gate errors depend on the phonon numbers tracked in Section~\ref{subsec:phonon_tracking}, while idle dephasing depends on the elapsed time experienced by each ion.

\subsubsection{Circuit Reconstruction}
The compiled instruction sequence is traversed in its canonical order to reconstruct the logical circuit in Stim format. Transport operations are omitted because they do not alter the ideal logical circuit, and their effects are instead incorporated through noise injection.

Each sequence from an ancilla's \texttt{Reset} to its corresponding \texttt{Measure} is reconstructed as a
stabilizer measurement. The MS gates are mapped to CNOT gates, with their operand order determining
whether the stabilizer is X- or Z-type. For X-type stabilizers, Hadamard gates are inserted before and
after the CNOT sequence. Native single-qubit rotations are not reconstructed individually, as their basis-change effects are represented by these Hadamard gates. Because scheduling may interleave instructions from different stabilizer measurements, reconstruction relies on the MS-gate structure rather than the original sequence of single-qubit rotations.

Native $R$ rotations are not reconstructed individually. They implement basis changes that fold into these Hadamards,
and scheduling may interleave them with independent instructions of other stabilizer blocks,
so the reconstruction relies only on the MS structure.

The reconstructed stabilizer measurements are assembled into a multi-round circuit in execution order. Detectors
compare corresponding ancilla measurements across consecutive rounds, and a final transversal data
measurement defines the logical observable. The reconstructed circuit is checked for agreement against the input circuit for the error-free case.

\subsubsection{Noise Injection}

Noise channels are inserted into the reconstructed Stim
circuit based on the compiled instruction sequence.
Two-qubit gate errors depend on the tracked phonon
numbers, while idle dephasing is determined from the
per-ion execution timelines. State preparation,
measurement, and single-qubit gate errors are modeled
using fixed error rates.

\paragraph{Two-Qubit Gate Errors}

After each CNOT gate, the pipeline inserts a general two-qubit Pauli channel with probabilities for the
15 non-identity Pauli errors. 
These probabilities are obtained from a precomputed lookup table in Ref.~\cite{hirai2026} derived from the master-equation simulation of an MS gate under thermal motional occupation, followed by Pauli twirling of the resulting process matrix.

The lookup table is indexed by the transverse mean
phonon number, corresponding to the motional direction
used for the MS gate. For each two-qubit gate, the
larger of the two participating ions' tracked values
is used to determine the error probabilities.

Since the lookup table is defined on a discrete grid,
the phonon number is rounded up to the next grid
point by default, providing a conservative error
estimate. Rounding down or to the nearest grid
point is also supported for sensitivity analysis.
Values exceeding the table's maximum are clipped
to its last entry and reported. The limitations
introduced by this clipping are discussed in Section~\ref{sec:conclusion}.

\paragraph{Idle Dephasing on the Per-Ion Timeline}
While one ion undergoes a gate or a transport, physical time passes for every other ion, during which its coherence decays. The pipeline reconstructs an absolute-time timeline for each ion from the compiled instruction sequence to account for dephasing during idle periods. An ion is considered \emph{busy} while participating in a computational operation and \emph{idle} otherwise, including during its own transport.
Before each reconstructed circuit operation, the accumulated idle time $\Delta t$ of each participating qubit since its previous busy interval is converted into a phase-flip error with probability
\begin{equation}
    p_Z(\Delta t) = \frac{1 - e^{-2\Gamma_\phi \Delta t}}{2},
    \label{eq:p_idle}
\end{equation}
where $\Gamma_\phi$ is the pure-dephasing rate.
Idle time preceding state preparation is discarded,
since phase errors immediately before a reset
have no observable effect.

\paragraph{Other Error Channels}
State preparation, measurement, and single-qubit gate errors are modeled using fixed error rates.
A bit-flip error is applied after each state preparation and before each measurement, while a depolarizing error is applied after each single-qubit gate. The corresponding error probabilities and other noise parameters are summarized in Appendix~\ref{app:setup}.

\subsubsection{Logical Error Rate Estimation}
Finally, the noisy Stim circuit is sampled to obtain detector outcomes and logical observable flips. For each shot,
PyMatching~~\cite{higgott2022pymatching} applies minimum-weight perfect matching (MWPM) to the detector outcomes to predict the logical observable flip.
The logical error rate $p_L$ is estimated as the fraction of shots for which the predicted logical observable flip disagrees with the sampled one.
\section{Impact of Cooling Strategies on QEC Performance}
\label{sec:experiments}

This section evaluates cooling strategies described in Section~\ref{subsec:cooling_policies} with three rounds of rotated-surface code syndrome extraction.
We judge logical error suppression with increasing code distance $\Lambda_{d,d+2} = p_L(d) / p_L(d+2)$.
The architecture parameters, noise channels, heating model, and modeling assumptions are shown in Appendix~\ref{app:setup}.

\subsection{Phonon-Number Requirement}
\label{subsec:exp_nbar_cap}

Before evaluating the cooling strategies, we determine how low the phonon number must be to achieve logical error suppression with increasing code distance. We compile each workload without cooling insertion and impose a virtual phonon-number ceiling $\bar{n}_{\mathrm{cap}}$ when evaluating gate errors. The phonon record evolves according to the compiled schedule, but values exceeding $\bar{n}_{\mathrm{cap}}$ are replaced by the ceiling when determining the gate error probabilities.

\begin{figure*}[htbp]
  \centering
  \includegraphics[width=\linewidth]{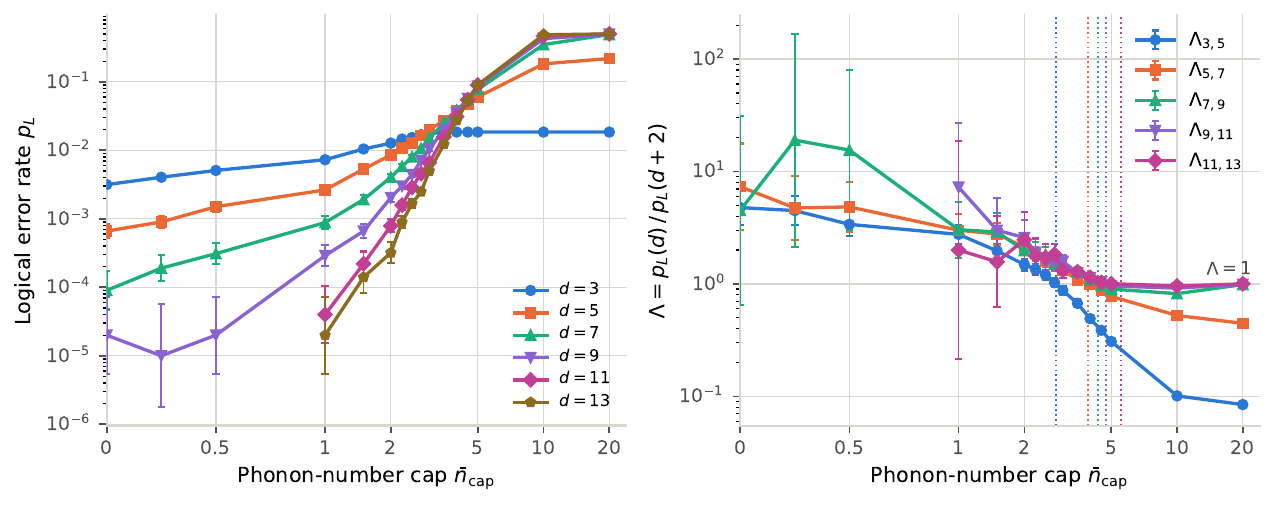}
  \caption{Logical error rate (left) and scaling ratio (right)
  versus the phonon-number ceiling $\bar{n}_{\mathrm{cap}}$
  on the no-cooling schedules over three rounds.
  Dotted vertical lines mark where each scaling ratio crosses one.
  This break-even ceiling rises with the code distance,
  from $\approx 2.8$ for $\Lambda_{3,5}$
  to $\approx 5.6$ for $\Lambda_{11,13}$.
  Points with no observed errors at $10^{5}$ shots are omitted.
  Ceilings of $20$ and above coincide with the uncapped register
  because the error-rate table saturates there.}
  \label{fig:nbar_cap}
\end{figure*}

Figure~\ref{fig:nbar_cap} shows the results. The logical error rate increases monotonically with $\bar{n}_{\mathrm{cap}}$ at every code distance. The scaling ratios fall below one at $\bar{n}_{\mathrm{cap}} \approx 2.8$ for $\Lambda_{3,5}$ and $\approx 3.9$ for $\Lambda_{5,7}$. The break-even ceiling increases for larger distance pairs, reaching approximately $5.6$ for $\Lambda_{11,13}$. Thus, among the tested pairs, $\Lambda_{3,5}$ imposes the strictest requirement. A ceiling of approximately $2.5$ yields logical error suppression for all tested distance pairs up to $d=13$. This establishes a reference phonon-number requirement against which the cooling strategies
are evaluated in the following subsections.

\subsection{Scaling without Cooling}
\label{subsec:exp_nocool}
Without cooling, increasing the code distance raises the logical error rate rather than suppressing it. The scaling ratio is $\Lambda_{3,5}=0.083$, and the logical error rate approaches $0.5$ at $d=7$ as shown in the first row of Table~\ref{tab:cooling_strategies}. This behavior is accompanied by increasing motional excitation. Larger codes require more transport operations and longer execution times, causing the average gate-time phonon number to rise from $1.6$ at $d=3$ to $6.3$ at $d=5$ and $20.8$ at $d=7$. These values substantially exceed the virtual ceiling of approximately $2.5$ identified in Section~\ref{subsec:exp_nbar_cap}. Thus, increasing the code distance also increases the physical noise experienced by the circuit, offsetting the additional error-correction capability.

To isolate the contribution of motional heating, we reevaluate the same compiled schedules with the phonon number used for every two-qubit gate error calculation fixed at zero. This removes the dependence of gate errors on the accumulated phonon record while preserving the execution schedule, idle dephasing, and other noise channels.

Under this control, the logical error rates decrease to $0.0032$, $0.00067$, and $0.00007$ for $d=3$, $5$, and $7$, respectively. The corresponding scaling ratios are $\Lambda_{3,5}=4.7$ and $\Lambda_{5,7}=9.6$.
Thus, under the modeled parameters, the loss of logical error suppression in the uncooled schedules is attributable to the heating-dependent gate errors rather than the remaining noise channels alone.

\subsection{Cooling Strategies Compared}
\label{subsec:exp_strategies}
We compare the cooling policies described in Section~\ref{subsec:cooling_policies} with $t_{\textrm{cool}}=1 \,\mathrm{ms}$ and $\bar{n}_{\textrm{res}}=0$. The effects of finite cooling duration and nonzero residual excitation are examined separately in Section~\ref{subsec:exp_requirements}.

The configurations differ in whether two coolant ions permanently reside in the LIZ and which additional cooling trigger is used. We consider baseline LIZ cooling alone, the memory predictive trigger with $\theta=1.5$ and $N_{\max}$ varied from $1$ to $16$, and the LIZ predictive trigger with $\theta$ varied at $N_{\max}=32$. The LIZ predictive trigger requires the resident coolant ions.

\begin{figure*}[t]
  \centering
  \includegraphics[width=0.9\linewidth]{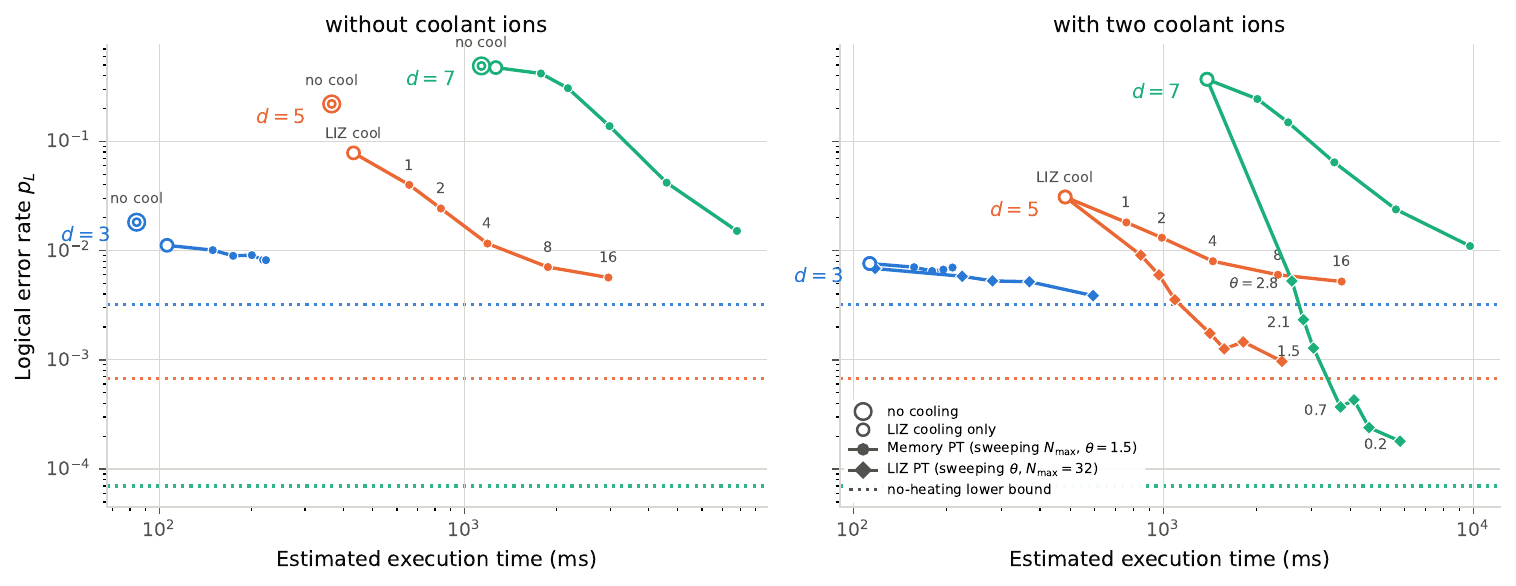}
  \caption{Logical error rate versus estimated execution time
  over three rounds,
  without (left) and with (right) two coolant ions in the LIZ.
  Double circles mark the uncooled baseline.
  Open circles mark LIZ cooling alone,
  filled circles the memory predictive trigger
  as $N_{\max}$ grows from $1$ to $16$ at $\theta = 1.5$,
  and diamonds the LIZ predictive trigger
  as $\theta$ falls from $2.8$ to $0.2$ at $N_{\max} = 32$.
  Dotted horizontal lines give the ideal case without heating, which is the achievable lower bound.}
  \label{fig:cooling_tradeoff}
\end{figure*}

Figure~\ref{fig:cooling_tradeoff} compares the logical error rates and estimated execution times of these configurations, while Table~\ref{tab:cooling_strategies} reports representative results.

\begin{table}[htbp]
  \caption{Representative cooling configurations
  (3 rounds, $10^{5}$ shots,
  $2 \times 10^{4}$ for the uncooled row).
  Starred $p_L$ values contain gates clipped
  at the lookup-table maximum and are lower bounds.
  MPT and LPT denote the memory and LIZ predictive triggers.\\}
  \label{tab:cooling_strategies}
  \centering
  \begin{tabular}{llllcc}
    \hline
    configuration & $p_L(3)$ & $p_L(5)$ & $p_L(7)$ & $\Lambda_{3,5}$ & $\Lambda_{5,7}$ \\
    \hline
    No cooling & $0.018$ & $0.220$ & $0.491^{*}$ & $0.08$ & $0.45$ \\
    LIZ cooling & $0.011$ & $0.078$ & $0.474^{*}$ & $0.14$ & $0.17$ \\
    \, + MPT ($N_{\max} = 16$) & $0.0082$ & $0.0057$ & $0.0151$ & $1.44$ & $0.37$ \\
    LIZ cooling + coolants & $0.0076$ & $0.031$ & $0.370^{*}$ & $0.25$ & $0.08$ \\
    \, + MPT ($N_{\max} = 16$) & $0.0070$ & $0.0052$ & $0.0110$ & $1.34$ & $0.47$ \\
    \, + LPT ($\theta = 0.2$) & $0.0039$ & $0.00097$ & $0.00018$ & $3.99$ & $5.39$ \\
    \hline
  \end{tabular}
\end{table}

Without resident coolant ions, LIZ cooling reduces the logical error rate compared to no cooling at every code distance but does not achieve logical error suppression with increasing distance. 

\begin{figure}[t]
  \centering
  \includegraphics[width=\linewidth]{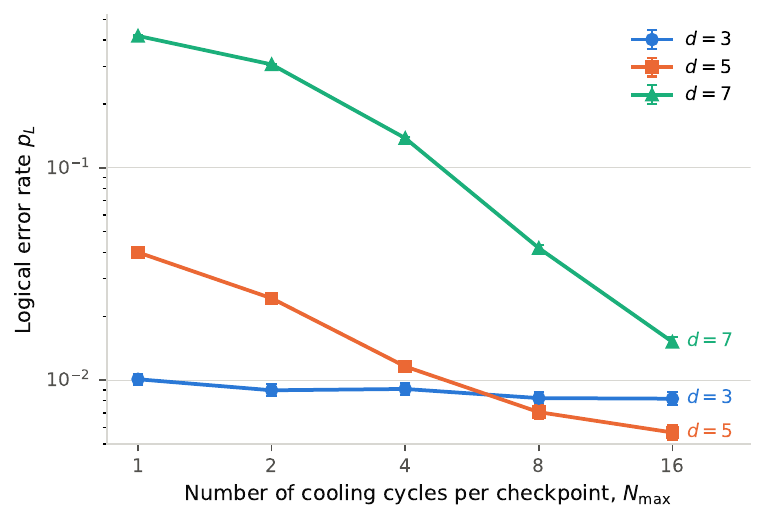}
  \caption{Logical error rate $p_{\textrm{L}}$ per three rounds versus the number of cooling cycles $N_{\max}$ per trigger point for the memory predictive trigger without coolant ions
  ($\theta = 1.5$).
  The $d = 5$ curve crosses below $d = 3$ near $N_{\max} = 8$,
  the point where $\Lambda_{3,5}$ exceeds one.}
  \label{fig:cooling_depth}
\end{figure}

Adding the memory predictive trigger improves the logical error rate as the maximum number of cooling cycles $N_{\max}$ increases, as shown in Fig.~\ref{fig:cooling_depth}. The scaling ratio $\Lambda_{3,5}$ first exceeds one at $N_{\max}=8$, reaching $1.16$ $[1.01,1.34]$, and increases to $1.44$ $[1.24,1.68]$ at $N_{\max}=16$. However, at $d=5$, the estimated execution time increases from $0.43\,\mathrm{s}$ with baseline LIZ cooling to $3.0\,\mathrm{s}$ at $N_{\max}=16$.

Although repeated cooling also reduces the logical error rate at $d=7$ by approximately a factor of 30, $\Lambda_{5,7}$ remains below one ($0.37$ at $N_{\max}=16$). Thus, the memory predictive trigger restores logical error suppression between $d=3$ and $d=5$, but not between $d=5$ and $d=7$, despite the additional cooling time.

The limitation of the memory predictive trigger arises from cooling an entire memory chain through phonon averaging (Eq.~\ref{eq:phonon_averaging}). When a cooled ancilla is merged with a chain of $k$ ions, the averaging step reduces the chain's excess phonon number by only a fraction $1/(k+1)$. Moreover, each merge introduces additional motional excitation, so repeated cooling approaches a nonzero phonon-number floor that increases with chain length.

This behavior is reflected in Figure~\ref{fig:cooling_depth}. At $d=3$, the logical error rate changes little beyond $N_{\max}=4$, whereas larger codes continue to benefit from additional cycles. Nevertheless, the longer memory chains at $d=7$ prevent the memory predictive trigger from achieving logical error suppression between $d=5$ and $d=7$ within the tested cycle budgets.

Adding two resident coolant ions to the LIZ reduces the logical error rate even without an additional cooling trigger. The coolant ions are cooled after each stabilizer measurement and subsequently participate in phonon averaging when data ions enter the LIZ. At $d=5$, this reduces the logical error rate under baseline LIZ cooling from $0.078$ to $0.031$.

The resident coolant ions also improve the performance of the memory predictive trigger, but do not eliminate its dependence on memory-chain length. At $N_{\max}=16$, the scaling ratios reach $\Lambda_{3,5}=1.34$ $[1.15,1.58]$ and $\Lambda_{5,7}=0.47$. Thus, although the additional coolant ions reduce the logical error rates, the memory predictive trigger still fails to achieve logical error suppression between $d=5$ and $d=7$ under the tested conditions.

In contrast, the LIZ predictive trigger cools the participating data--ancilla pair immediately before each two-qubit gate when the tracked phonon number exceeds $\theta$. By isolating the pair and averaging its motional excitation with the resident coolant ions, this policy acts on a small group of ions rather than an entire memory chain. Its cooling efficiency therefore does not decrease with memory-chain length in the same way as that of the memory predictive trigger.

At $\theta=0.2$ and $N_{\max}=32$, the logical error rates are $0.0039$, $0.00097$, and $0.00018$ for $d=3$, $5$, and $7$, respectively. The corresponding suppression ratios are $\Lambda_{3,5}=3.99$ $[2.96,5.37]$ and $\Lambda_{5,7}=5.39$ $[2.80,10.4]$, with both 95\% confidence intervals entirely above one. This is the first configuration in our comparison to achieve logical error suppression across both distance pairs.

At $d=7$, the logical error rate is within a factor of $2.6$ of the no-heating reference established in Section~\ref{subsec:exp_nocool}, with an estimated execution time of $5.8\,\mathrm{s}$ for three syndrome-extraction rounds. At both $d=5$ and $d=7$, the LIZ predictive trigger also achieves lower logical error rates with shorter execution times than any of the tested memory predictive trigger configurations.

We close this comparison with a methodological observation.
Evaluating trigger-based policies requires the time-proportional background heating of the model, not only the per-operation costs.
With per-operation heating only, the memory predictive trigger at $N_{\max} = 1$ measures $p_L = 0.0328$ $[0.0317, 0.0339]$ at $d = 5$.
Adding the background heating moves it to $0.0400$ $[0.0388, 0.0412]$, a relative change of $22\%$
with disjoint intervals, and raises the number of fired cycles
from $180$ to $239$.
A policy operating at its threshold boundary is sensitive to the quanta that accumulate between its trigger points, in both its cost and its benefit, and a static or per-operation heating estimate
misses this effect entirely.

\subsection{Sweeping the Gate-Time Phonon Ceiling}
\label{subsec:exp_theta}

The threshold of the LIZ predictive trigger
acts as a direct control of the phonon number
at gate execution.
With a sufficient cycle budget and the stopping conditions of
Section~\ref{subsec:cooling_policies},
the trigger cools before every two-qubit gate
whose data qubit exceeds $\theta$,
and the measured maximum gate-time $\bar{n}$
follows $\theta$ within $0.04$ quanta
across the sweep.
Sweeping $\theta$ therefore re-runs the thought experiment of
Section~\ref{subsec:exp_nbar_cap}
with a real policy that pays real time.

\begin{figure}[t]
  \centering
  \includegraphics[width=\linewidth]{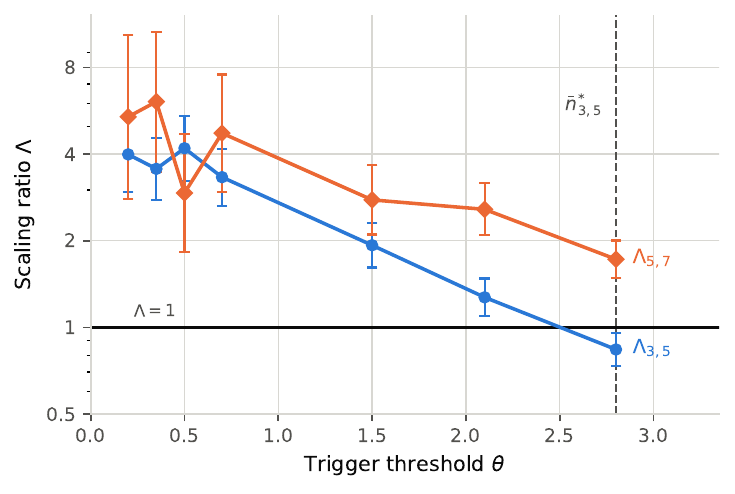}
  \caption{Scaling ratios versus the trigger threshold $\theta$
  for the LIZ predictive trigger
  ($N_{\max} = 32$,
  idealized cooling at $1\,\mathrm{ms}$ to zero residual).
  Error bars are endpoint ratios of the Wilson 95\% intervals.
  The dashed vertical line marks the break-even ceiling
  $\bar{n}^{*}_{3,5} \approx 2.8$
  measured independently in Figure~\ref{fig:nbar_cap}.}
  \label{fig:lpt_theta}
\end{figure}

Figure~\ref{fig:lpt_theta} shows the two ratios
across $\theta \in [0.2, 2.8]$.
$\Lambda_{3,5}$ crosses one between
$\theta = 2.1$ ($1.27$ $[1.09, 1.48]$)
and $\theta = 2.8$ ($0.84$ $[0.73, 0.96]$),
in agreement with the break-even ceiling
$\bar{n}^{*}_{3,5} \approx 2.8$
obtained on the uncooled schedules.
The two experiments measure the same boundary
by independent means,
one by capping the record without paying time
and one by paying the time of a trigger policy.
The threshold itself promises no bound at gate time,
because the cooling episode is followed by
the merge back and the transport to the gate.
Measured, the maximum gate-time phonon number exceeds $\theta$ by at most $0.05$ quanta across the sweep,
so the $\theta$ axis is comparable to the virtual ceiling,
and the two boundaries agree on where scaling begins.
$\Lambda_{5,7}$ exceeds one over the whole sweep,
from $1.72$ $[1.48, 2.0]$ at $\theta = 2.8$.

Below $\theta \approx 0.7$ the ratios flatten
within their intervals
while the execution time keeps growing
($1.4$ to $2.4\,\mathrm{s}$ at $d = 5$
from $\theta = 0.7$ to $0.2$),
and the $d = 3$ error rate continues to improve
as the dilution approaches its floor.
We adopt $\theta = 0.2$ as the canonical setting
for the requirement study below,
the coldest point of the sweep.

\subsection{Cooling Requirements for Scalable QEC}
\label{subsec:exp_requirements}

The results so far used an idealized cooling instruction with $t_{\textrm{cool}}=1\,\mathrm{ms}$ and $\bar{n}_{\textrm{res}}=0$.
The cooling methods in Section~\ref{subsec:cooling} differ in their attainable motional temperatures and cooling durations,
so a physical realization of the instruction corresponds to a finite duration $t_{\mathrm{cool}}$ and a nonzero residual $\bar{n}_{\mathrm{res}}$.
To measure the required performance of the cooling method for the LIZ predictive trigger policy, we re-run the canonical configuration of Section~\ref{subsec:exp_theta} with every cooling instruction replaced by a duration and residual pair, over $t_{\mathrm{cool}} = 1$ to $75\,\mathrm{ms}$ and $\bar{n}_{\mathrm{res}} = 0$ to $6$.

The trigger threshold follows the residual as
$\theta = \max(0.2,\ \bar{n}_{\mathrm{res}} + 0.1)$,
since a threshold below the residual
would fire without effect.
At every sampled positive residual,
this $\theta$ lies below the fixed point of the cooling cycle,
the residual plus the split and merge costs
of the isolation round trip
($0.15$ quanta in our parameters),
so the threshold cannot be reached,
and each triggered episode ends at the floor-detection
condition of Section~\ref{subsec:cooling_policies},
cooling the gate pair toward that floor.
At zero residual the threshold $0.2$
sits above the fixed point,
and each episode ends by reaching it.
In both regimes the episode terminates itself,
well before the cycle budget $N_{\max}$.
Across the grid,
the measured maximum gate-time phonon number
exceeds the residual by an amount
that grows with the duration,
from $0.24$ quanta at $t_{\mathrm{cool}} = 1\,\mathrm{ms}$
to $0.9$ at $10$
and $6.3$ at $75\,\mathrm{ms}$.

\begin{figure*}[t]
  \centering
  \includegraphics[width=0.9\linewidth]{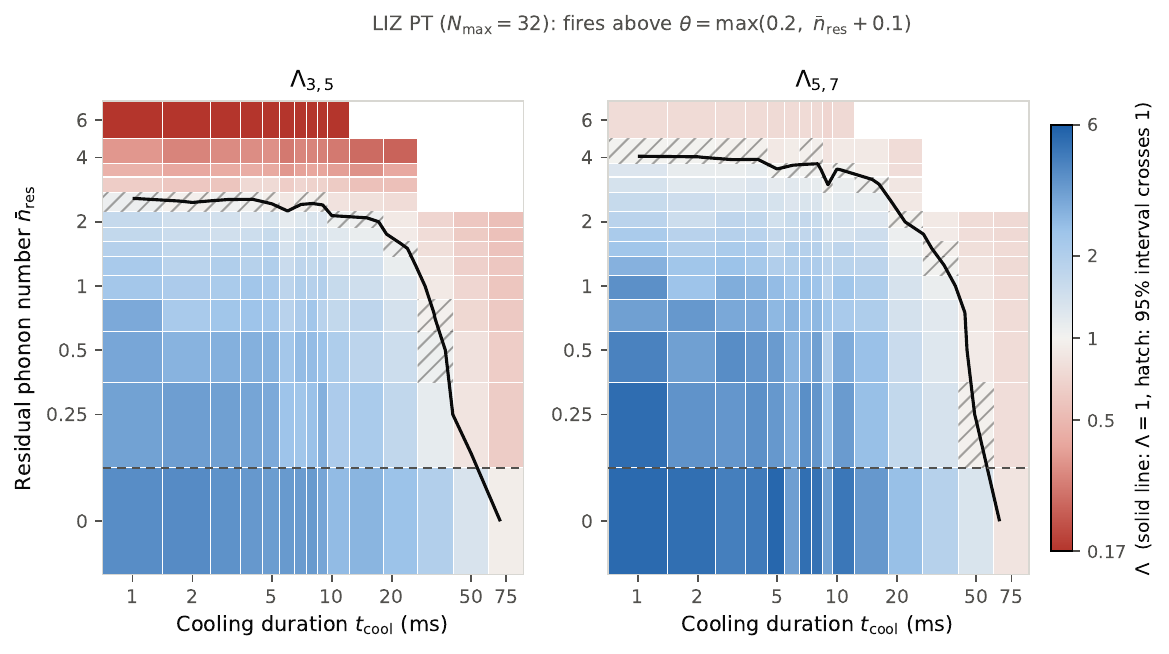}
  \caption{Error suppression ratios (color) of the LIZ predictive trigger ($N_{\max} = 32$) with each cooling instruction assigned duration $t_{\mathrm{cool}}$ and post-cooling residual phonon number $\bar{n}_{\mathrm{res}}$.
  The predictive trigger fires above $\theta = \max(0.2,\ \bar{n}_{\mathrm{res}} + 0.1)$.
  The solid curve is the $\Lambda = 1$ contour, hatching marks cells whose 95\% intervals cross one, and white cells far beyond the boundary are not simulated.
  The row below the dashed line is $\bar{n}_{\mathrm{res}} = 0$,
  placed on a linear segment of the axis.}
  \label{fig:req_grid}
\end{figure*}

Figure~\ref{fig:req_grid} shows that,
for durations of one to ten milliseconds,
the LIZ predictive trigger satisfies
the scaling criterion
in $89$ of the $90$ tested combinations
with $\bar{n}_{\mathrm{res}} \leq 2$.
$\Lambda_{5,7}$ satisfies the criterion
throughout this range (minimum $1.36$).
The only exception for $\Lambda_{3,5}$
is at $t_{\mathrm{cool}} = 10\,\mathrm{ms}$
and $\bar{n}_{\mathrm{res}} = 2.0$,
where the point estimate is $1.07$
but its 95\% interval includes one.
At $t_{\mathrm{cool}} = 1\,\mathrm{ms}$,
raising the residual from $0$ to $2$
lowers $\Lambda_{3,5}$ from $3.99$ to $1.62$,
whereas at zero residual,
increasing the cooling duration
from $1$ to $10\,\mathrm{ms}$
lowers it from $3.99$ to $2.64$.
The scaling ratio is therefore more sensitive
to the residual phonon number
in these comparisons.

Beyond the low-residual, short-duration region,
the scaling boundary of each pair shifts
toward smaller residual phonon numbers
as the cooling duration increases,
following the growth of the measured gate-time excess.
At short durations,
where that excess is about $0.25$ quanta,
the residual boundaries fall where the break-even ceilings
of Section~\ref{subsec:exp_nbar_cap} predict.
The transition for $\Lambda_{3,5}$ occurs
between $\bar{n}_{\mathrm{res}} = 2.5$ and $3.0$,
close to the ceiling-based estimate of $2.6$,
and the transition for $\Lambda_{5,7}$
between $3.5$ and $4.0$,
close to the estimate of $3.7$.
Even at zero residual,
both scaling ratios satisfy the criterion
at $50\,\mathrm{ms}$
but fall below one at $75\,\mathrm{ms}$
($\Lambda_{3,5} = 1.34$ and $\Lambda_{5,7} = 1.31$,
then $0.93$ and $0.85$).

The resulting map characterizes the combinations
of cooling duration and residual phonon number
under which the LIZ predictive trigger supports
logical-error scaling in the evaluated 1D QCCD model.
These requirements concern the parameters of
the abstract cooling instruction,
rather than the performance of
any particular physical cooling method.
\section{Conclusion}
\label{sec:conclusion}

We presented a transport- and heating-aware evaluation pipeline for quantum error correction with explicit cooling strategies on a 1D QCCD architecture. It includes a compiler that translates syndrome measurement circuits with transport and cooling operations, tracks the mean phonon number of every ion through each instruction it emits, and feeds the tracked state through gate-dependent error channels into the logical error rate.

The experiments of Section~\ref{sec:experiments}
turn this pipeline into three findings.
First, without cooling, enlarging the code distance of surface codes fails to suppress logical error rates.
Comparison with simulations without heating shows that this is caused by heating.
Second, within the evaluated 1D QCCD model, the choice of cooling strategy determines whether logical error rates decrease across both tested code-distance pairs.
Strategies that cool through the memory chains (\emph{memory predictive trigger}) are limited by a chain-length-dependent phonon floor, whereas the \emph{LIZ predictive trigger}, which cools the gate pair against coolant ions immediately before each two-qubit gate, achieves scaling ratios above one for both pairs ($\Lambda_{3,5} = 3.99$ and $\Lambda_{5,7} = 5.39$), with their 95\% intervals entirely above one.
Third, the LIZ predictive trigger satisfies the scaling criterion
for nearly all tested combinations with cooling durations of one to ten milliseconds and residual phonon numbers up to two. The extended sweep reveals a downward-sloping boundary in the duration and residual plane. The two cooling parameters trade against each other.
At short durations, the residual boundaries are consistent with the virtual-ceiling analysis, and even at zero residual, scaling is observed at $50\,\mathrm{ms}$ but not at $75\,\mathrm{ms}$.
These results identify the required performance for the abstract cooling instruction under the evaluated LIZ predictive trigger strategy.

This evaluation has several limitations.
Compilation and cooling strategies at larger distances and rounds, with different QCCD architectures, and the performance of other code families remain future work.

The phonon record is idealized.
Every ion in a chain carries the same mean phonon number per mode, merging equalizes chains by the phonon averaging of
Section~\ref{subsec:cooling}, and a cooling instruction sets its targets to a residual value after a fixed duration.
Real cooling methods enter only through that duration and residual pair, not through their method-specific dynamics.
The assumed phonon averaging mechanism does not resolve the individual motional modes or the finite-time dynamics of energy exchange.
Consequently, the simulated benefit of phonon-average indirect cooling is conditional on this idealization.
The per-operation phonon costs of splitting, merging,
and reordering are fixed parameters
whose hardware-specific calibration remains open.

The instruction durations and the background heating rates
are taken from particular experimental reports
and do not represent every platform.
Our conclusions are statements about this parameter point, although the required study of
Section~\ref{sec:experiments} spans the cooling parameters themselves.

\section*{Acknowledgement}
EI defined the instruction set and intermediate representation, implemented the compilation flow and simulator backend, and performed numerical calculations and data visualization.
NH performed numerical calculations of the  M{\o}lmer--S{\o}rensen gate infidelity as a function of phonon number, which were required for deriving the logical error rate. Details of this work are available in Ref.~\cite{hirai2026}.
KM supervised EI in defining the instruction set and guided practical constraints of QCCD hardware. KM also provided guidance on the numerical analyses based on experience operating actual QCCD hardware.
HE contributed to the basic strategy for the compilation process, including qubit allocation for the 1D QCCD architecture, and performed calculations for the code cycle times with boustrophedon allocation.
KT co-supervised NH and provided advice on the numerical methods.
IK assisted NH with the numerical calculations.
KT contributed to the discussion on the heating model and the cooling strategies.
MT supervised NH and provided advice on the numerical methods and the quantities to be investigated.
HT contributed to discussions on compilation and cooling strategies for QCCD devices based on experience operating actual hardware.
TS supervised EI’s undergraduate thesis research, which forms the basis of the compilation flow presented in the first part of this work.
SN designed the compilation and simulation flow and the cooling strategy, implemented numerical calculations for determining the phonon number threshold, made some figures, and supervised the overall project.

\bibliographystyle{ieeetr}
\bibliography{eitaroLiterature}

@unpublished{hirai2026,
  author = {Hirai, N. and Tanji, K. and Miyanishi, K. and Nishio, S. and Satoh, T. and Takahashi, H. and Takeoka, M.},
  title  = {Analysis of Robustness of Mølmer-Sørensen gates under Thermal Motion for Circuit-level Noise Simulation},
  year   = {2026},
  note   = {Manuscript in preparation}
}

@article{Ye2025quantumerror,
  doi = {10.22331/q-2025-11-27-1920},
  url = {https://doi.org/10.22331/q-2025-11-27-1920},
  title = {Quantum error correction for long chains of trapped ions},
  author = {Ye, Min and Delfosse, Nicolas},
  journal = {{Quantum}},
  issn = {2521-327X},
  publisher = {{Verein zur F{\"{o}}rderung des Open Access Publizierens in den Quantenwissenschaften}},
  volume = {9},
  pages = {1920},
  month = nov,
  year = {2025}
}

@book{nielsen2010quantum,
  author    = {Nielsen, Michael A. and Chuang, Isaac L.},
  title     = {Quantum Computation and Quantum Information},
  edition   = {10th Anniversary},
  publisher = {Cambridge University Press},
  year      = {2010},
  isbn      = {978-1-107-00217-3}
}

@article{Gidney_2021,
   title={Stim: a fast stabilizer circuit simulator},
   volume={5},
   ISSN={2521-327X},
   url={http://dx.doi.org/10.22331/q-2021-07-06-497},
   DOI={10.22331/q-2021-07-06-497},
   journal={Quantum},
   publisher={Verein zur Forderung des Open Access Publizierens in den Quantenwissenschaften},
   author={Gidney, Craig},
   year={2021},
   month=jul, pages={497} 
}

@article{HAFFNER_2008,
   title={Quantum computing with trapped ions},
   volume={469},
   ISSN={0370-1573},
   url={http://dx.doi.org/10.1016/j.physrep.2008.09.003},
   DOI={10.1016/j.physrep.2008.09.003},
   number={4},
   journal={Physics Reports},
   publisher={Elsevier BV},
   author={H{\"a}ffner, H. and Roos, C. F. and Blatt, R.},
   year={2008},
   month=dec, pages={155--203}
}

@article{M_lmer_1999,
   title={Multiparticle Entanglement of Hot Trapped Ions},
   volume={82},
   ISSN={1079-7114},
   url={http://dx.doi.org/10.1103/PhysRevLett.82.1835},
   DOI={10.1103/physrevlett.82.1835},
   number={9},
   journal={Physical Review Letters},
   publisher={American Physical Society (APS)},
   author={Mølmer, Klaus and Sørensen, Anders},
   year={1999},
   month=mar, pages={1835--1838} 
}

@article{kielpinski2002architecture,
  title={Architecture for a large-scale ion-trap quantum computer},
  author={Kielpinski, David and Monroe, Chris and Wineland, David J},
  journal={Nature},
  volume={417},
  number={6890},
  pages={709--711},
  year={2002},
  publisher={Nature Publishing Group}
}

@inproceedings{jones2025architectingscalabletrappedion,
author = {Jones, Scott and Murali, Prakash},
title = {Architecting Scalable Trapped Ion Quantum Computers using Surface Codes},
year = {2026},
isbn = {9798400723599},
publisher = {Association for Computing Machinery},
address = {New York, NY, USA},
url = {https://doi.org/10.1145/3779212.3790128},
doi = {10.1145/3779212.3790128},
booktitle = {Proceedings of the 31st ACM International Conference on Architectural Support for Programming Languages and Operating Systems, Volume 2},
pages = {175--190},
numpages = {16},
location = {Pittsburgh, PA, USA},
series = {ASPLOS '26}
}

@article{schoenberger2024shuttlingscalabletrappedionquantum,
author = {Schoenberger, Daniel and Hillmich, Stefan and Brandl, Matthias and Wille, Robert},
title = {Shuttling for Scalable Trapped-Ion Quantum Computers},
year = {2025},
issue_date = {June 2025},
publisher = {IEEE Press},
volume = {44},
number = {6},
issn = {0278-0070},
url = {https://doi.org/10.1109/TCAD.2024.3513262},
doi = {10.1109/TCAD.2024.3513262},
journal = {Trans. Comp.-Aided Des. Integ. Cir. Sys.},
month = jun,
pages = {2144--2155},
numpages = {12}
}

@inproceedings{murali2020architectingnoisyintermediatescaletrapped,
  title={Architecting noisy intermediate-scale trapped ion quantum computers},
  author={Murali, Prakash and Debroy, Dripto M and Brown, Kenneth R and Martonosi, Margaret},
  booktitle={2020 ACM/IEEE 47th Annual International Symposium on Computer Architecture (ISCA)},
  pages={529--542},
  year={2020},
  organization={IEEE}
}

@article{Maslov_2017,
   title={Basic circuit compilation techniques for an ion-trap quantum machine},
   volume={19},
   ISSN={1367-2630},
   url={http://dx.doi.org/10.1088/1367-2630/aa5e47},
   DOI={10.1088/1367-2630/aa5e47},
   number={2},
   journal={New Journal of Physics},
   publisher={IOP Publishing},
   author={Maslov, Dmitri},
   year={2017},
   month=feb, pages={023035} }

@article{Poschinger_2009,
   title={Coherent manipulation of a40Ca+spin qubit in a micro ion trap},
   volume={42},
   ISSN={1361-6455},
   url={http://dx.doi.org/10.1088/0953-4075/42/15/154013},
   DOI={10.1088/0953-4075/42/15/154013},
   number={15},
   journal={Journal of Physics B: Atomic, Molecular and Optical Physics},
   publisher={IOP Publishing},
   author={Poschinger, U G and Huber, G and Ziesel, F and Deiß, M and Hettrich, M and Schulz, S A and Singer, K and Poulsen, G and Drewsen, M and Hendricks, R J and Schmidt-Kaler, F},
   year={2009},
   month=jul, pages={154013} }

@article{hilder2021faulttolerantparityreadoutshuttlingbased,
  title = {Fault-Tolerant Parity Readout on a Shuttling-Based Trapped-Ion Quantum Computer},
  author = {Hilder, J. and Pijn, D. and Onishchenko, O. and Stahl, A. and Orth, M. and Lekitsch, B. and Rodriguez-Blanco, A. and M\"uller, M. and Schmidt-Kaler, F. and Poschinger, U. G.},
  journal = {Phys. Rev. X},
  volume = {12},
  issue = {1},
  pages = {011032},
  numpages = {16},
  year = {2022},
  month = {Feb},
  publisher = {American Physical Society},
  doi = {10.1103/PhysRevX.12.011032},
  url = {https://link.aps.org/doi/10.1103/PhysRevX.12.011032}
}

@article{kaushal2019shuttlingbasedtrappedionquantuminformation,
  title={Shuttling-based trapped-ion quantum information processing},
  author={Kaushal, Vidyut and Lekitsch, Bjoern and Stahl, A and Hilder, J and Pijn, Daniel and Schmiegelow, C and Bermudez, Alejandro and M{\"u}ller, M and Schmidt-Kaler, Ferdinand and Poschinger, U},
  journal={AVS Quantum Science},
  volume={2},
  number={1},
  year={2020},
  publisher={AIP Publishing}
}

@article{shor1995scheme,
  title={Scheme for reducing decoherence in quantum computer memory},
  author={Shor, Peter W},
  journal={Physical review A},
  volume={52},
  number={4},
  pages={R2493},
  year={1995},
  publisher={APS}
}

@misc{lark,
  author = {Erez Shinan},
  title = {Lark - A parsing toolkit for Python},
  year = {2025},
  publisher = {GitHub},
  journal = {GitHub repository},
  howpublished = {\url{https://github.com/lark-parser/lark}},
  note         = {Accessed: 2026-01-29}
}

@misc{iso14977,
  author       = {{ISO/IEC}},
  title        = {{ISO/IEC 14977:1996 Information technology -- Syntactic metalanguage -- Extended BNF}},
  howpublished = {International Organization for Standardization},
  year         = {1996},
  address      = {Geneva, Switzerland},
  url          = {https://www.iso.org/standard/26153.html},
  note         = {Accessed: 2026-01-29}
}

@article{Dennis_2002,
   title={Topological quantum memory},
   volume={43},
   ISSN={1089-7658},
   url={http://dx.doi.org/10.1063/1.1499754},
   DOI={10.1063/1.1499754},
   number={9},
   journal={Journal of Mathematical Physics},
   publisher={AIP Publishing},
   author={Dennis, Eric and Kitaev, Alexei and Landahl, Andrew and Preskill, John},
   year={2002},
   month=sep, pages={4452--4505} }

@article{Bruzewicz_2019,
   title={Trapped-ion quantum computing: Progress and challenges},
   volume={6},
   ISSN={1931-9401},
   url={http://dx.doi.org/10.1063/1.5088164},
   DOI={10.1063/1.5088164},
   number={2},
   journal={Applied Physics Reviews},
   publisher={AIP Publishing},
   author={Bruzewicz, Colin D. and Chiaverini, John and McConnell, Robert and Sage, Jeremy M.},
   year={2019},
   month=may }

@article{Fowler_2012,
   title={Surface codes: Towards practical large-scale quantum computation},
   volume={86},
   ISSN={1094-1622},
   url={http://dx.doi.org/10.1103/PhysRevA.86.032324},
   DOI={10.1103/physreva.86.032324},
   number={3},
   journal={Physical Review A},
   publisher={American Physical Society (APS)},
   author={Fowler, Austin G. and Mariantoni, Matteo and Martinis, John M. and Cleland, Andrew N.},
   year={2012},
   month=sep }

@article{Tomita_2014,
   title={Low-distance surface codes under realistic quantum noise},
   volume={90},
   ISSN={1094-1622},
   url={http://dx.doi.org/10.1103/PhysRevA.90.062320},
   DOI={10.1103/physreva.90.062320},
   number={6},
   journal={Physical Review A},
   publisher={American Physical Society (APS)},
   author={Tomita, Yu and Svore, Krysta M.},
   year={2014},
   month=dec }

@book{gottesman1997stabilizercodesquantumerror,
  title={Stabilizer codes and quantum error correction},
  author={Gottesman, Daniel},
  year={1997},
  publisher={California Institute of Technology}
}

@article{Scowen1993GenericBS,
  title={Generic base standards},
  author={Roger S. Scowen},
  journal={Proceedings 1993 Software Engineering Standards Symposium},
  year={1993},
  pages={25-34},
  url={https://api.semanticscholar.org/CorpusID:17097372}
}

@article{Harty_2014,
  title = {High-Fidelity Preparation, Gates, Memory, and Readout of a Trapped-Ion Quantum Bit},
  author = {Harty, T. P. and Allcock, D. T. C. and Ballance, C. J. and Guidoni, L. and Janacek, H. A. and Linke, N. M. and Stacey, D. N. and Lucas, D. M.},
  journal = {Phys. Rev. Lett.},
  volume = {113},
  pages = {220501},
  year = {2014},
  publisher = {American Physical Society},
  doi = {10.1103/PhysRevLett.113.220501}
}

@article{Ballance_2016,
  title = {High-Fidelity Quantum Logic Gates Using Trapped-Ion Hyperfine Qubits},
  author = {Ballance, C. J. and Harty, T. P. and Linke, N. M. and Sepiol, M. A. and Lucas, D. M.},
  journal = {Phys. Rev. Lett.},
  volume = {117},
  pages = {060504},
  year = {2016},
  publisher = {American Physical Society},
  doi = {10.1103/PhysRevLett.117.060504}
}

@article{Gaebler_2016,
  title = {High-Fidelity Universal Gate Set for ${^{9}\mathrm{Be}}^{+}$ Ion Qubits},
  author = {Gaebler, J. P. and Tan, T. R. and Lin, Y. and Wan, Y. and Bowler, R. and Leibfried, D. and Wineland, D. J.},
  journal = {Phys. Rev. Lett.},
  volume = {117},
  pages = {060505},
  year = {2016},
  publisher = {American Physical Society},
  doi = {10.1103/PhysRevLett.117.060505}
}

@inproceedings{10.1145/3624062.3624214,
author = {Leblond, Tyler and Bennink, Ryan S. and Lietz, Justin G. and Seck, Christopher M.},
title = {TISCC: A Surface Code Compiler and Resource Estimator for Trapped-Ion Processors},
year = {2023},
isbn = {9798400707858},
publisher = {Association for Computing Machinery},
address = {New York, NY, USA},
url = {https://doi.org/10.1145/3624062.3624214},
doi = {10.1145/3624062.3624214},
booktitle = {Proceedings of the SC '23 Workshops of the International Conference on High Performance Computing, Network, Storage, and Analysis},
pages = {1426--1435},
numpages = {10},
location = {Denver, CO, USA},
series = {SC-W '23}
}

@article{higgott2022pymatching,
  author    = {Higgott, Oscar},
  title     = {{PyMatching}: A {Python} Package for Decoding Quantum Codes with Minimum-Weight Perfect Matching},
  journal   = {ACM Transactions on Quantum Computing},
  volume    = {3},
  number    = {3},
  articleno = {16},
  year      = {2022},
  doi       = {10.1145/3505637}
}

@article{Kreppel_2023,
  author  = {Kreppel, Fabian and Melzer, Christian and Olvera Mill{\'a}n, Diego and Wagner, Janis and Hilder, Janine and Poschinger, Ulrich and Schmidt-Kaler, Ferdinand and Brinkmann, Andr{\'e}},
  title   = {Quantum Circuit Compiler for a Shuttling-Based Trapped-Ion Quantum Computer},
  journal = {Quantum},
  volume  = {7},
  pages   = {1176},
  year    = {2023},
  doi     = {10.22331/q-2023-11-08-1176}
}

@article{pino2021demonstration,
  author  = {Pino, Juan M. and Dreiling, Jennifer M. and Figgatt, Caroline and Gaebler, John P. and Moses, Steven A. and Allman, M. S. and Baldwin, Charles H. and Foss-Feig, Michael and Hayes, David and Mayer, Karl and Ryan-Anderson, Ciar{\'a}n and Neyenhuis, Brian},
  title   = {Demonstration of the trapped-ion quantum {CCD} computer architecture},
  journal = {Nature},
  volume  = {592},
  pages   = {209--213},
  year    = {2021},
  doi     = {10.1038/s41586-021-03318-4}
}

@inproceedings{saki2022muzzle,
  author    = {Saki, Abdullah Ash and Topaloglu, Rasit Onur and Ghosh, Swaroop},
  title     = {Muzzle the Shuttle: Efficient Compilation for Multi-Trap Trapped-Ion Quantum Computers},
  booktitle = {Proceedings of the Design, Automation \& Test in Europe Conference \& Exhibition (DATE)},
  year      = {2022}
}

@article{eschner2003laser,
  title = {Laser cooling of trapped ions},
  author = {Eschner, J{\"u}rgen and Morigi, Giovanna and Schmidt-Kaler, Ferdinand and Blatt, Rainer},
  journal = {J. Opt. Soc. Am. B},
  volume = {20},
  number = {5},
  pages = {1003--1015},
  year = {2003},
  publisher = {Optica Publishing Group},
  doi = {10.1364/JOSAB.20.001003}
}

@article{diedrich1989zeropoint,
  title = {Laser Cooling to the Zero-Point Energy of Motion},
  author = {Diedrich, F. and Bergquist, J. C. and Itano, Wayne M. and Wineland, D. J.},
  journal = {Phys. Rev. Lett.},
  volume = {62},
  issue = {4},
  pages = {403--406},
  year = {1989},
  publisher = {American Physical Society},
  doi = {10.1103/PhysRevLett.62.403},
  url = {https://link.aps.org/doi/10.1103/PhysRevLett.62.403}
}

@article{roos2000eit,
  title = {Experimental Demonstration of Ground State Laser Cooling with Electromagnetically Induced Transparency},
  author = {Roos, C. F. and Leibfried, D. and Mundt, A. and Schmidt-Kaler, F. and Eschner, J. and Blatt, R.},
  journal = {Phys. Rev. Lett.},
  volume = {85},
  issue = {26},
  pages = {5547--5550},
  year = {2000},
  publisher = {American Physical Society},
  doi = {10.1103/PhysRevLett.85.5547},
  url = {https://link.aps.org/doi/10.1103/PhysRevLett.85.5547}
}

@article{larson1986sympathetic,
  title = {Sympathetic cooling of trapped ions: A laser-cooled two-species nonneutral ion plasma},
  author = {Larson, D. J. and Bergquist, J. C. and Bollinger, J. J. and Itano, Wayne M. and Wineland, D. J.},
  journal = {Phys. Rev. Lett.},
  volume = {57},
  issue = {1},
  pages = {70--73},
  year = {1986},
  publisher = {American Physical Society},
  doi = {10.1103/PhysRevLett.57.70},
  url = {https://link.aps.org/doi/10.1103/PhysRevLett.57.70}
}

@article{fallek2024exchange,
  title = {Rapid exchange cooling with trapped ions},
  author = {Fallek, Spencer D. and Sandhu, Vikram S. and McGill, Ryan A. and Gray, John M. and Tinkey, Holly N. and Clark, Craig R. and Brown, Kenton R.},
  journal = {Nature Communications},
  volume = {15},
  pages = {1089},
  year = {2024},
  publisher = {Nature Publishing Group},
  doi = {10.1038/s41467-024-45232-z},
  url = {https://www.nature.com/articles/s41467-024-45232-z}
}

@article{sorensen2000entanglement,
  title = {Entanglement and quantum computation with ions in thermal motion},
  author = {S{\o}rensen, Anders and M{\o}lmer, Klaus},
  journal = {Phys. Rev. A},
  volume = {62},
  issue = {2},
  pages = {022311},
  year = {2000},
  publisher = {American Physical Society},
  doi = {10.1103/PhysRevA.62.022311},
  url = {https://link.aps.org/doi/10.1103/PhysRevA.62.022311}
}

@article{ejtemaee2017sisyphus,
  title = {3D Sisyphus Cooling of Trapped Ions},
  author = {Ejtemaee, S. and Haljan, P. C.},
  journal = {Phys. Rev. Lett.},
  volume = {119},
  issue = {4},
  pages = {043001},
  year = {2017},
  publisher = {American Physical Society},
  doi = {10.1103/PhysRevLett.119.043001},
  url = {https://link.aps.org/doi/10.1103/PhysRevLett.119.043001}
}

@article{roos1999sideband,
  title = {Quantum State Engineering on an Optical Transition and Decoherence in a Paul Trap},
  author = {Roos, Ch. and Zeiger, Th. and Rohde, H. and N\"agerl, H. C. and Eschner, J. and Leibfried, D. and Schmidt-Kaler, F. and Blatt, R.},
  journal = {Phys. Rev. Lett.},
  volume = {83},
  issue = {23},
  pages = {4713--4716},
  year = {1999},
  publisher = {American Physical Society},
  doi = {10.1103/PhysRevLett.83.4713}
}

@article{barrett2003sympathetic,
  title = {Sympathetic cooling of {$^{9}$Be$^{+}$} and {$^{24}$Mg$^{+}$} for quantum logic},
  author = {Barrett, M. D. and DeMarco, B. and Schaetz, T. and Leibfried, D. and Britton, J. and Chiaverini, J. and Itano, W. M. and Jelenkovic, B. M. and Jost, J. D. and Langer, C. and Rosenband, T. and Wineland, D. J.},
  journal = {Phys. Rev. A},
  volume = {68},
  pages = {042302},
  year = {2003},
  doi = {10.1103/PhysRevA.68.042302},
  url = {https://doi.org/10.1103/PhysRevA.68.042302}
}

@article{home2009memory,
  title = {Memory coherence of a sympathetically cooled trapped-ion qubit},
  author = {Home, J. P. and McDonnell, M. J. and Szwer, D. J. and Keitch, B. C. and Lucas, D. M. and Stacey, D. N. and Steane, A. M.},
  journal = {Phys. Rev. A},
  volume = {79},
  pages = {050305},
  year = {2009},
  doi = {10.1103/PhysRevA.79.050305},
  url = {https://doi.org/10.1103/PhysRevA.79.050305}
}

@article{allcock2021omg,
  title = {{omg} blueprint for trapped ion quantum computing with metastable states},
  author = {Allcock, D. T. C. and Campbell, W. C. and Chiaverini, J. and Chuang, I. L. and Hudson, E. R. and Moore, I. D. and Ransford, A. and Roman, C. and Sage, J. M. and Wineland, D. J.},
  journal = {Applied Physics Letters},
  volume = {119},
  pages = {214002},
  year = {2021},
  doi = {10.1063/5.0069544},
  url = {https://doi.org/10.1063/5.0069544}
}

@article{yang2022dualtype,
  title = {Realizing coherently convertible dual-type qubits with the same ion species},
  author = {Yang, H.-X. and Ma, J.-Y. and Wu, Y.-K. and Wang, Y. and Cao, M.-M. and Guo, W.-X. and Huang, Y.-Y. and Feng, L. and Zhou, Z.-C. and Duan, L.-M.},
  journal = {Nature Physics},
  volume = {18},
  pages = {1058--1061},
  year = {2022},
  doi = {10.1038/s41567-022-01661-5},
  url = {https://doi.org/10.1038/s41567-022-01661-5}
}

@article{wineland1987lasercooling,
  title = {Laser-cooling limits and single-ion spectroscopy},
  author = {Wineland, D. J. and Itano, W. M. and Bergquist, J. C. and Hulet, R. G.},
  journal = {Phys. Rev. A},
  volume = {36},
  number = {5},
  pages = {2220--2232},
  year = {1987},
  doi = {10.1103/PhysRevA.36.2220},
  url = {https://doi.org/10.1103/PhysRevA.36.2220}
}

@misc{jones2026wiser,
  title = {{WISER}: Systematic Design-Space Exploration of Trapped Ions
           with Multiplexed Control},
  author = {Jones, Scott and Yang, Song-qing-hao and Murali, Prakash},
  year = {2026},
  eprint = {2609.13546},
  archivePrefix = {arXiv},
  primaryClass = {quant-ph},
}

@article{smith2025singlequbit,
  title = {Single-qubit gates with errors at the $10^{-7}$ level},
  author = {Smith, M. C. and Leu, A. D. and Miyanishi, K. and Gely, M. F. and Lucas, D. M.},
  journal = {Physical Review Letters},
  volume = {134},
  pages = {230601},
  year = {2025},
  doi = {10.1103/42w2-6ccy}
}

@misc{hughes2025smoothgate,
  title = {Trapped-ion two-qubit gates with $>$99.99\% fidelity without ground-state cooling},
  author = {Hughes, A. C. and Srinivas, R. and L{\"o}schnauer, C. M. and Knaack, H. M. and Matt, R. and Ballance, C. J. and Malinowski, M. and Harty, T. P. and Sutherland, R. T.},
  year = {2025},
  eprint = {2510.17286},
  archivePrefix = {arXiv},
  primaryClass = {quant-ph}
}

@inproceedings{ovide2024scaling,
  title={Scaling and assigning resources on ion trap qccd architectures},
  author={Ovide, Anabel and Cuomo, Daniele and Almudever, Carmen G},
  booktitle={2024 IEEE international conference on Quantum Computing and Engineering (QCE)},
  volume={1},
  pages={959--970},
  year={2024},
  organization={IEEE}
}

@misc{theata_github,
  author       = {Eitaro Ishikawa and Shin Nishio},
  title        = {{THeatA}: Transport- and Heating-aware
                  QEC Simulation Framework},
  year         = {2026},
  howpublished = {\url{https://github.com/MSTakahashiPJ/theata}},
  note         = {GitHub repository}
}

\appendices
\section{Experimental Setup}
\label{app:setup}

\paragraph{Architecture and circuits}
All experiments use the single-LIZ zone configuration \texttt{M-M-M-L-M-M-M} introduced in Section~\ref{subsec:qccd} with LIZ capacity $4$. Each memory zone has sufficient capacity to hold at least half of the ions in the register, with additional space available. The memory and temporary zone capacities are set so that they do not constrain the compiled schedules. We confirmed that reducing these capacities does not change the generated instruction sequences.

Each workload is $r = 3$ rounds of rotated surface code syndrome extraction circuit.
Each round is compiled separately with the stabilizer measurement order alternating between forward and reverse order. The final ion placement of each round is used as the initial placement for the next round.
The CNOT order within each stabilizer follows the hook-error-avoiding schedule ~\cite{Tomita_2014}, and the initial placement alternates qubits between the two memory zones, balancing the register around the central LIZ. More details on initial placements are discussed in Appendix~\ref{app:qubit-mapping}.

\paragraph{Noise channels and decoding}
The compiled schedule is converted to a Stim circuit and decoded with minimum-weight perfect matching decoder as described in Section~\ref{subsec:noise_pipeline}.
Two-qubit gate errors are derived per gate from the phonon-dependent lookup table calculated in Ref.~\cite{hirai2026},
keyed by the transverse phonon number of the hotter participating ion and rounded upward (the conservative mode).
The table covers $\bar{n} \le 20$.
Gates above this range are clipped to the table maximum, so the reported logical error rate is a lower bound
wherever clipping occurs, and we flag every such point.
Idle qubits accumulate dephasing at $\Gamma_\phi = 3.0\,\mathrm{s}^{-1}$ over their per-ion idle gaps.

The remaining channels use
$p_R = 10^{-4}$,
$p_\mathrm{meas} = 3.2 \times 10^{-3}$,
and $p_\mathrm{reset} = 8 \times 10^{-4}$.
Each data point uses $2 \times 10^{4}$ decoded shots,
and the points that decide a scaling ratio
use $10^{5}$ shots.
Uncertainties are Wilson $1\sigma$ intervals.

\paragraph{Heating model}
Each ion carries three motional accumulators,
one axial and two transverse
(Section~\ref{subsec:phonon_tracking}).
Transport increments use the following defaults:
shuttling adds $0.5$ axial quanta per $180\,\mu\mathrm{m}$
and $0.05$ transverse quanta per operation,
splitting and merging add $10$ axial
and $0.05$ transverse quanta,
with the merge cost applied after
the chain-wide averaging,
and an in-chain swap adds $0.1$ quanta in every direction.
On top of these per-operation costs,
ions heat continuously with elapsed time
at the background rates measured by
Kaushal et al.~\cite{kaushal2019shuttlingbasedtrappedionquantuminformation},
$5.5$ quanta/s axial and
$4.8$ and $3.6$ quanta/s
for the two transverse modes.
The worse transverse rate is assigned to the mode
that the gate error model reads.
Elapsed time is accounted on a planning-time clock,
the serial sum of emitted instruction durations,
which is an upper bound on the scheduled execution time
and therefore conservative.
Cooling resets every accumulator of the addressed chain
to a residual value of zero
at a time cost of $400\,\mu\mathrm{s}$,
and the deep re-cooling of measurement ancillae
uses $2000\,\mu\mathrm{s}$.

\paragraph{Modeling assumptions}
Three simplifications bound the interpretation of the results.
First, phonon accumulators reset at round boundaries, so background heating accumulates within a round only,
and its reported effect is a lower bound.
Second, a cooling instruction occupies only its own zone. 
Transport in other zones may proceed in parallel, which matches the zone parallelism of the QCCD architecture,
and we verified on every compiled schedule that no measurement or gate overlaps
a cooling cycle in time.
Third, the phonon-averaging model of Section~\ref{subsec:cooling}
assumes complete equalization of the mean phonon numbers
across the merged chain.

\section{Intermediate Representation}
\label{app:ir}

This appendix gives the full grammar of the
intermediate representation introduced in
Section~\ref{subsec:translation},
together with the semantics of each instruction.
The DSL is specified in extended Backus--Naur 
form~\cite{iso14977, Scowen1993GenericBS}
and parsed with Lark~\cite{lark},
whose grammar dialect the listing below uses verbatim.
In this DSL,
a program is a sequence of qubit declarations,
which fix the initial placement,
followed by instructions.
Listing~\ref{lst:lark_grammar} gives the grammar in full.

\begin{lstlisting}[caption={Complete Lark grammar of the QCCD compiler's DSL.}, label={lst:lark_grammar}, basicstyle=\ttfamily\footnotesize]  
    start: (declaration | instruction)+

    // Declarations (initial placement)
    declaration: "QUBIT" ion_id "AT" location

    // Instructions
    instruction: gate_op | movement_op | measure_op
               | reset_op | barrier_op | cool_op
               | comment

    // Basic types
    ion_id:    INT
    trap_id:   INT
    site_id:   INT
    angle:     SIGNED_NUMBER
    location:  "(" trap_id "," site_id ")"
    zone_id:   INT

    // Gate operations
    gate_op: ms_op | r_op
    ms_op: "MS" ion_id ion_id angle
    r_op:  "R"  ion_id angle angle

    // Movement operations
    movement_op: shuttle_op | split_op
               | merge_op | swap_op

    shuttle_op: "SHUTTLE" ion_id "->" location
    swap_op:    "SWAP" ion_id ion_id
    split_op:   "SPLIT" ion_id "->" zone_id
    merge_op:   "MERGE" zone_id zone_id

    // Other operations
    measure_op:  "MEASURE" ion_id+
    reset_op:    "RESET"   ion_id+
    barrier_op:  "BARRIER" ion_id+

    // Cooling
    cool_op:     "COOL" zone_id ion_id?

    comment: /#.*/

    // Terminal definitions (Lark)
    %import common.INT
    %import common.SIGNED_NUMBER
    %import common.WS
    %import common.NEWLINE
    %ignore WS
    %ignore NEWLINE
\end{lstlisting}

\paragraph{Instruction set}
Operands written $i$ and $j$ are ion identifiers,
$z$ is a zone identifier,
and $(t, s)$ is a site $s$ of trap $t$.

\newcommand{\irop}[2]{\par\smallskip\noindent #1\par\nopagebreak
  {\leftskip 1em\relax\noindent #2\par}}

\irop{\texttt{QUBIT} $i$ \texttt{AT} $(t, s)$}{%
  declares ion $i$ with initial location $(t, s)$.}
\irop{\texttt{MS} $i$ $j$ $\theta$}{%
  applies a M{\o}lmer--S{\o}rensen entangling
  gate~\cite{M_lmer_1999}
  with angle $\theta$ to ions $i$ and $j$,
  which must sit in the same interaction zone.}
\irop{\texttt{R} $i$ $\theta$ $\phi$}{%
  applies a single-qubit rotation
  with angle $\theta$ and phase $\phi$ to ion $i$.}
\irop{\texttt{SHUTTLE} $i$ \texttt{->} $(t, s)$}{%
  transports ion $i$ to $(t, s)$.
  Only the destination is written.
  The source follows from the placement
  that the compiler tracks.}
\irop{\texttt{SWAP} $i$ $j$}{%
  exchanges the order of ions $i$ and $j$,
  which must reside in the same zone.}
\irop{\texttt{SPLIT} $i$ \texttt{->} $z$}{%
  cuts the chain holding ion $i$
  immediately to the right of $i$,
  so that $i$ becomes the right end of its segment,
  and moves the segment on the side of $z$ into zone $z$.
  The moving side follows from the direction
  from the source zone to $z$ along the trap axis.}
\irop{\texttt{MERGE} $z_{l}$ $z_{r}$}{%
  joins the chains of the adjacent zones
  $z_{l}$ and $z_{r}$,
  preserving the physical left-to-right order.}
\irop{\texttt{MEASURE} $i^{+}$ / \texttt{RESET} $i^{+}$}{%
  measures, or resets, the listed ions
  in the interaction zone.}
\irop{\texttt{BARRIER} $i^{+}$}{%
  is a synchronization point for the listed ions
  and takes no time.
  It pins the relative order of instructions
  that would otherwise be independent.}
\irop{\texttt{COOL} $z$ $[c]$}{%
  cools the entire chain held in zone $z$,
  resetting each of its motional accumulators
  to the residual value of the configured cooling method.
  The optional operand $c$ names a coolant ion
  that must be present in $z$ when the instruction executes,
  which the resident-coolant configurations use.
  A per-instruction duration distinguishes
  the regular cooling
  from the deep post-measurement re-cooling
  (Appendix~\ref{app:setup}).}
\medskip

Gate, transport, and cooling instructions
are charged the durations and heating increments
of the execution-time estimator of
Section~\ref{subsec:estimator},
with the parameter values of Appendix~\ref{app:setup}.

\paragraph{Dependency-graph construction}
Algorithm~\ref{alg:dag_construction} constructs
the dependency graph of Section~\ref{subsec:dag}
in a single pass over the instruction sequence,
maintaining a lookup table \texttt{last\_op}
that records the most recent instruction
touching each qubit.

\begin{algorithm}[htbp]
    \caption{Dependency Graph Construction}
    \label{alg:dag_construction}
    \begin{algorithmic}[1]
        \Require Sequence of computational instructions $I_{seq}$ parsed from the AST
        \Ensure Directed acyclic graph $G = (V, E)$
        \State $V \leftarrow \emptyset,\ E \leftarrow \emptyset$
        \State $\texttt{last\_op} \leftarrow \text{empty map: qubit ID} \to \text{node ID}$
        \For{\textbf{each} instruction $I_{new}$ \textbf{in} $I_{seq}$}
            \State Create node $v_{new}$ for $I_{new}$;\ \ $V \leftarrow V \cup \{v_{new}\}$
            \For{\textbf{each} qubit $q$ used by $I_{new}$}
                \If{$q \in \texttt{last\_op}$}
                    \State Add edge $(\texttt{last\_op}[q], v_{new})$ to $E$
                \EndIf
                \State $\texttt{last\_op}[q] \leftarrow v_{new}$
            \EndFor
        \EndFor
        \State \Return $G(V, E)$
    \end{algorithmic}
\end{algorithm}

\section{Qubit Mapping for Time-efficient Syndrome Extraction}
\label{app:qubit-mapping}

The initial ion placement affects the transport schedule and code cycle time for syndrome extraction in a one-dimensional QCCD architecture. It has also been studied as a means of reducing ion transport overhead in QCCD architectures. Ovide et al.~\cite{ovide2024scaling} proposed a spatio-temporal-aware allocation heuristic that exploits qubit interactions and their temporal ordering to reduce execution time of NISQ algorithms. In contrast, our placement strategy exploits the geometric structure of rotated surface-code stabilizers and the memory–LIZ separation of our linear architecture. In this appendix, we optimize the initial ion placement for a syndrome-extraction circuit that reuses a single ancilla qubit to measure multiple stabilizer checks. This ancilla-reuse syndrome extraction was studied
in~\cite{Ye2025quantumerror}.

\begin{figure}[htb]
 \centering
 \includegraphics[width=\linewidth]{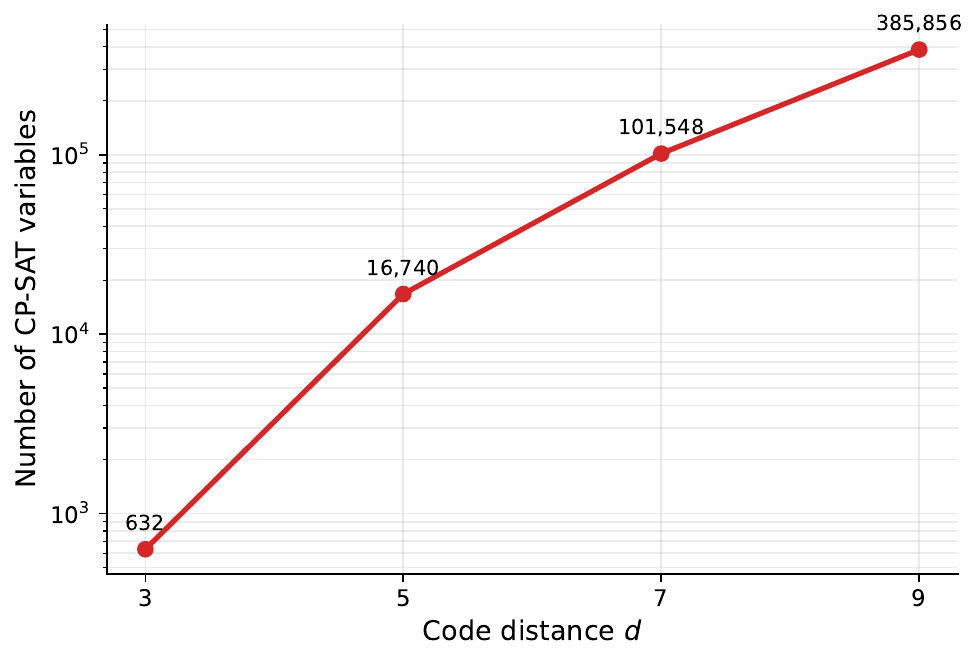}
 \caption{Number of variables in the time-expanded joint CP-SAT model. 
 Model size depends on both code distance and the maximum number of SWAP steps, which was set using a feasible solution obtained before CP-SAT optimization.
 }
 \label{fig:joint-cpsat-scaling}
\end{figure}

In principle, one could jointly optimize the initial order, gate schedule, and transport operations. We tested this approach using a time-expanded CP-SAT model for a reuse-one circuit on an abstract chain. For $d=3$, a 600-s run produced a replay-validated 16-SWAP incumbent and a lower bound of 10, so the optimum lies in $[10,16]$ but was not identified exactly. The upper--lower gap grew rapidly at larger distances, and with a 120-s limit CP-SAT did not establish a validated feasible incumbent for $d\geq7$, even though independently replayable warm-start trajectories were available. The time-expanded formulation therefore scales poorly because its state and transition variables grow with both the number of ions and the SWAP horizon. Figure~\ref{fig:joint-cpsat-scaling} shows the resulting model size.
The number of variables increases from 632 at $d=3$ to 385,856 at $d=9$, an
increase by more than two orders of magnitude. 

This limitation motivates a low-cost placement strategy based on code geometry. We first use \emph{stabilizer span} to identify a useful one-dimensional order and then adapt that order to the two memory zones surrounding the LIZ configuration. 

\subsection{From span minimization to boustrophedon placement}

For a data-qubit placement $\pi$ and a stabilizer code $C$ with stabilizer group $\mathscr{S} = \langle S_1, S_2, \dots, S_{n-k} \rangle$, we define \emph{stabilizer span} for a stabilizer check $S_i$ and \emph{code span} as 
\begin{equation}
 \begin{aligned}
S_{i, \mathrm{span}}(\pi)
   &\triangleq \max_{q\in\operatorname{supp}(S_i)}\pi(q)
      - \min_{q\in\operatorname{supp}(S_i)}\pi(q), \\
 C_{\mathrm{span}}(\pi)
   &\triangleq \sum_{S_i\in\mathscr{S}}S_{i,\mathrm{span}}(\pi)
 \end{aligned}
 \label{eq:total-span}
\end{equation}
where $\textrm{supp}(S_i)$ is a support of the stabilizer $S_i$.
The stabilizer span measures how far apart the data qubits of each stabilizer are in the initial ion ordering.
This locality is relevant to the transport cost required for syndrome measurement.

We minimized $C_{\mathrm{span}}(\pi)$ with CP-SAT by assigning every data
qubit a distinct integer position. For the $d=3$ rotated surface code,
the solver proved $C_{\mathrm{span}}=20$ optimal. The resulting placement
required 27 SWAPs under a simple router, compared with 40 for
sequential with original qubit index and 102 for a random placement. 
Span is therefore a useful placement proxy on this small instance to minimize the number of operations required for transport, although it does not fully determine the QCCD transport schedule.

\begin{figure}[htbp]
    \centering
    \begin{subfigure}{0.60\linewidth}
        \centering
        \includegraphics[width=\linewidth]{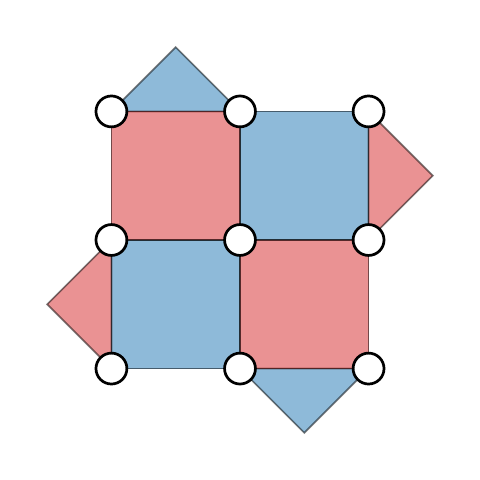}
        \caption{}
        \label{fig:stab}
    \end{subfigure}
    \hfill
    \begin{subfigure}{0.39\linewidth}
        \centering
        \includegraphics[width=\linewidth]{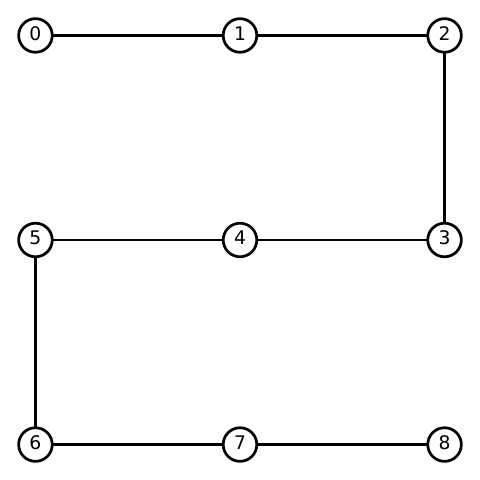}
        \caption{}
        \label{fig:boustrophedon}
    \end{subfigure}    
    \caption{(a) $d=3$ rotated surface code. Vertices are data qubits, and faces are stabilizer checks. (b) Boustrophedon index of data qubits in (a).}
    \label{fig:app_layout}
\end{figure}

\begin{figure}[htbp]
    \centering
    \includegraphics[width=\linewidth]{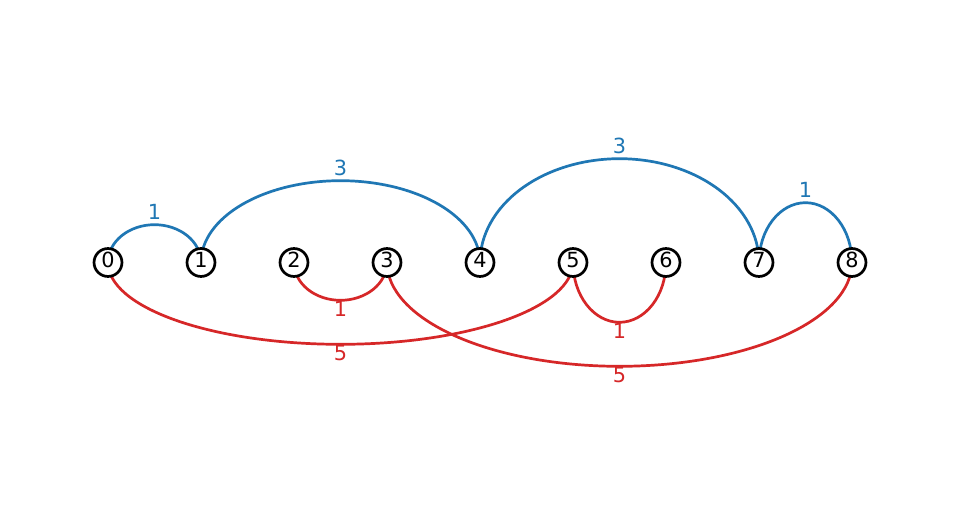}
    \caption{1D boustrophedon placement for $d=3$ rotated surface code. 
    Data qubits are shown as vertices, and X/Z stabilizer spans are shown as red/blue edges. $C_{\mathrm{span}}= 1 + 3 + 3 + 1 + 5 + 1 + 5 + 1 = 20$ and its optimality has been proven by CP-SAT.}
    \label{fig:app_span_on_chain}
\end{figure}

The CP-SAT output followed a clear geometric pattern. The first lattice row is scanned left-to-right, the next right-to-left, and the direction alternates thereafter, as an example shown in Fig.~\ref{fig:app_layout}.  Fig.~\ref{fig:app_span_on_chain} shows its stabilizer spans.
We call this the \emph{boustrophedon} order.
Table~\ref{tab:span-layout} shows that it matches the certified optimum at $d=3$ and stays close to the 60-s CP-SAT incumbents at larger distances. Those larger incumbents are not certified optima.

\begin{table}[t]
 \centering
 \caption{Total span of the CP-SAT incumbent and boustrophedon placement.\\}
 \label{tab:span-layout}
 \begin{tabular}{c|r|r}
  \hline
  $d$ & CP-SAT & Boustrophedon \\
  \hline
  3 & 20 & 20 \\
  5 & 103 & 104 \\
  7 & 288 & 300 \\
  9 & 638 & 656 \\
  \hline
 \end{tabular}
\end{table}

\subsection{Placement comparison}

We compare four representative placement strategies for one syndrome-extraction round of $d\in\{3,5,7,9\}$ rotated surface codes.
Sequential ordering of qubits by original qubit index, 
uniformly shuffled random order, minimizing code span with 30,000 simulated-annealing steps,
and Boustrophedon directly follows the alternating row traversal.
For each placement, CX gates for a stabilizer check are ordered by the leftmost to the rightmost positions of their support. 
Hook-safe gate order within a stabilizer is retained.
Figure~\ref{fig:placement-only} reports the QCCD compiler's
SWAP count and execution time. Boustrophedon has the lowest SWAP count at all
tested distances and the shortest simulated time among these four methods.

\subsection{Architecture-aware Alternating Side Placement}
The span-based placement treats the ions as a single chain, whereas our QCCD architecture has two memory zones on opposite sides of the LIZ. We therefore introduce \emph{alternating-side placement}.

To construct the syndrome-extraction schedule from a given initial data-ion placement, we first determine the order of stabilizer measurements. For each stabilizer, we identify the minimum and maximum positions of its data ions in the given one-dimensional placement (e.g. boustrophedon placement). Stabilizers are then ordered lexicographically by these two positions, prioritizing those whose leftmost data ion appears earlier in the chain. Within each stabilizer measurement, the CNOT gates follow a fixed order (NZ order) chosen to avoid hook errors.

Given this schedule, we scan the gates in execution order and record the order in which ions first appear. The ions are then rearranged alternately into the left and right memory zones, one ion at a time. Within each memory zone, ions appearing earlier are placed closer to the LIZ. Figure~\ref{fig:alternating-side-placement} shows an example of alternating side placement for 9 data ions. The original one-dimensional placement thus affects the alternating-side assignment through the circuit order, rather than through a direct mapping of ion positions. At last, the compiler inserts the required transport operations.

\setcounter{figure}{12}
\begin{figure}[htbp]
 \centering
 \includegraphics[width=\linewidth]{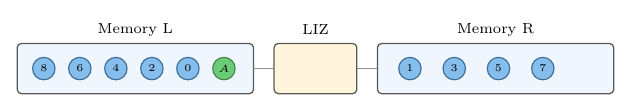}
 \caption{
 Alternating-side placement for nine data ions (blue) and one auxiliary qubit (green). Assuming that the ions first appear in the circuit in the order $(0,1,\ldots,8)$, they are assigned alternately to the left and right memory zones. Within each zone, ions appearing earlier are placed closer to the LIZ. The measurement ancilla is at the LIZ-facing end of the left memory zone. }
 \label{fig:alternating-side-placement}
\end{figure}

\setcounter{figure}{11}
\begin{figure*}[htbp]
 \centering
 \includegraphics[width=\linewidth]{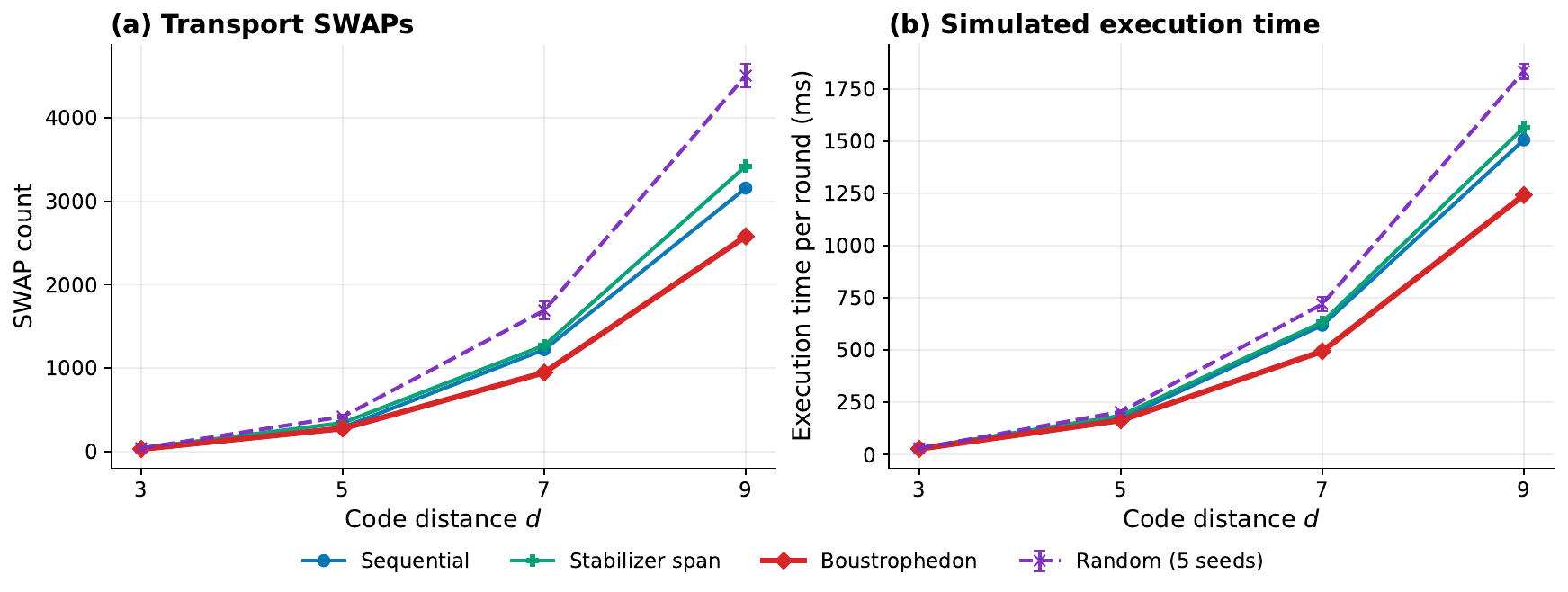}
 \caption{One-round QCCD cost for four initial-placement methods. Random error bars show one standard deviation over five seeds. At $d=9$, Boustrophedon requires 2580 SWAPs and 1243.2~ms, compared with the five-seed Random mean of 4508.2 SWAPs and 1835.0~ms.}
 \label{fig:placement-only}
\end{figure*}

Figure~\ref{fig:alternating-side-lines} compares the same four placements after this alternating side placement. 
For $d\geq5$, Boustrophedon plus alternating side placement has
the lowest SWAP count and execution time. At $d=9$, it requires 2018 SWAPs
and 893.0~ms. Relative to plain Boustrophedon, Alternating side placement reduces its SWAP count by 25.5\%, 19.9\%, and 21.8\% for $d=5,7,9$, respectively. 
At $d=3$, the reduction from 29 to 28 SWAPs is accompanied by a small time increase from 27.55 to 28.98~ms, illustrating that SWAP count is not an exact proxy for execution time.

These results identify two complementary effects. Boustrophedon provides a
low-cost geometric order that keeps stabilizer supports compact, while
alternating side placement maps the resulting circuit to both LIZ-facing
memory ends. 

\setcounter{figure}{13}
\begin{figure*}[htbp]
 \centering
 \includegraphics[width=\linewidth]{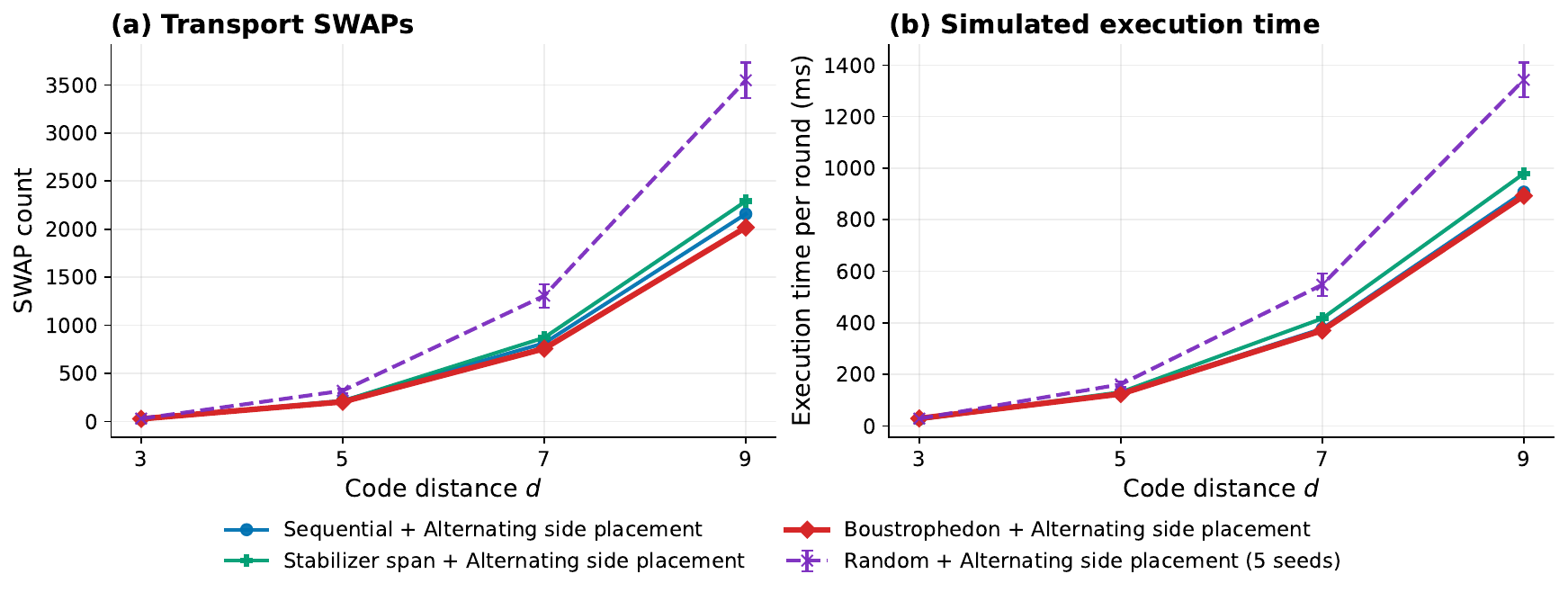}
 \caption{One-round QCCD cost after alternating side placement is applied to the circuit
 order derived from each source placement.}
 \label{fig:alternating-side-lines}
\end{figure*}

\EOD
\end{document}